\documentclass[conference]{IEEEtran}
\IEEEoverridecommandlockouts

\usepackage[final]{pdfpages}

\usepackage{mathtools}
\usepackage{marvosym}
\let\marvosymLightning\Lightning
\usepackage{wasysym}

\usepackage{stmaryrd}

\usepackage{listings}
\usepackage{subcaption}
\usepackage[inkscapelatex=false]{svg}
\usepackage{multicol}
\usepackage[noend,linesnumbered,ruled]{algorithm2e}
\usepackage{fancyvrb}
\SetAlFnt{\small}
\usepackage{graphicx}
\usepackage{textcomp}
\usepackage{cite}
\usepackage{amsmath,amssymb,amsfonts}
\usepackage{booktabs}
\usepackage{url}
\usepackage{tikz}
\usepackage{xcolor,colortbl}
\usepackage{hyperref}

\usepackage{caption}
\usepackage[dvipsnames]{xcolor}

\definecolor{darkyellow}{RGB}{255,168,67}
\definecolor{darkgreen}{RGB}{85,130,54}
\definecolor{lightgray}{RGB}{192,192,192}

\definecolor{addressOfColor}{rgb}{0.85,0.1529,0.1568}
\definecolor{assignColor}{rgb}{0.58039,0.40392,0.74117}
\definecolor{loadColor}{rgb}{0.54901,0.33725,0.29411}
\definecolor{storeColor}{rgb}{0.89019,0.46666,0.76078}

\definecolor{pointToColor}{rgb}{0.1215,0.4666,0.70588}
\definecolor{pointToDeltaColor}{rgb}{0.1725,0.62745,0.1725}
\definecolor{pointToNewColor}{rgb}{1,0.5,0.0549}

\newcommand{\smalltt}[1]{\texttt{\small #1}}

\definecolor{maroon}{cmyk}{0,0.87,0.68,0.32}

\newcommand{\blue}[1]{#1}

\newcommand{\Rnew}{$\text{\smalltt{R}}_{\text{\smalltt{new}}}$}
\newcommand{\Rdelta}{$\text{\smalltt{R}}_{\text{\smalltt{delta}}}$}
\newcommand{\Rbase}{$\text{\smalltt{R}}$}

\newcommand{\RnewInCode}{$\text{R}_{\text{new}}$}

\newcommand{\RbaseInCode}{$\text{R}$}

\newcommand{\Anew}{$\text{\smalltt{A}}_{\text{\smalltt{new}}}$}
\newcommand{\Adelta}{$\text{\smalltt{A}}_{\text{\smalltt{delta}}}$}
\newcommand{\Abase}{$\text{\smalltt{A}}$}

\newcommand{\AnewInCode}{$\text{A}_{\text{new}}$}
\newcommand{\AdeltaInCode}{$\text{A}_{\text{delta}}$}
\newcommand{\AbaseInCode}{$\text{A}$}

\newcommand{\AndersensAnalysis}{Andersen's analysis}

\newcolumntype{L}[1]{>{\raggedright\let\newline\\\arraybackslash\hspace{0pt}}m{#1}}
\newcolumntype{C}[1]{>{\centering\let\newline\\\arraybackslash\hspace{0pt}}m{#1}}
\newcolumntype{R}[1]{>{\raggedleft\let\newline\\\arraybackslash\hspace{0pt}}m{#1}}

\makeatletter
\newcommand{\removelatexerror}{\let\@latex@error\@gobble}
\renewcommand*{\@algocf@post@ruled}{}
\makeatother

\newcommand{\engine}{PlayLog}
\newcommand{\CI}{\textbf{CI}}
\newcommand{\UKI}{\textbf{UKI}}
\newcommand{\UPI}{\textbf{UPI}}
\newcommand{\FS}{\textbf{FS}}

\newcommand{\ARRAY}{\textbf{SA}}
\newcommand{\ARRAYALTERNATIVE}{\ARRAY{}\textbf{++}}
\newcommand{\BP}{\textbf{BP}}
\newcommand{\RX}{\textbf{RX}}
\newcommand{\HT}{\textbf{HT}}
\newcommand{\BL}{\textbf{RS}}

\newcommand{\DEDUPOne}{\textbf{S1}}
\newcommand{\DEDUPTwo}{\textbf{S2}}
\newcommand{\DEDUPThree}{\textbf{S3}}
\newcommand{\DEDUPFour}{\textbf{S4}}

\SetKwFor{ForPar}{for}{do in parallel}{end for}
\SetKw{KwAnd}{and}
\SetKw{KwAlloc}{allocate}

\lstdefinestyle{cppstyle}{
  language=C++,
  frame=tb,
  aboveskip=2mm,
  belowskip=2mm,
  captionpos=b,
  showstringspaces=false,
  columns=flexible,
  basicstyle={\scriptsize\ttfamily},
  numbers=left,
  numberstyle=\scriptsize \color{black},
  keywordstyle=\color{blue},
  commentstyle=\color{magenta},
  frame=none,
  breaklines=true,
  breakatwhitespace=true,
  tabsize=3,
  morekeywords=[2]{M_CREATE, M_BULK_INSERT, M_GET_NUMBER_OF_TUPLES, M_CONTAINS, M_APPEND, M_APPEND_FINISHED, M_ITERATOR_ALL, M_ITERATOR_RANGE, M_MOVE},
  keywordstyle=[2]\color{orange},  
  morekeywords=[3]{R_RECURSIVE_RULE, R_NON_RECURSIVE_RULE, R_COPY},
  keywordstyle=[3]\color{darkgreen},  
}
\newcommand{\refReviewS}[2]{\noindent\textbf{\hyperref[R#1S#2]{R#1S#2}}}
\newcommand{\refReviewO}[2]{\noindent\textbf{\hyperref[R#1O#2]{R#1O#2}}}
\newcommand{\refReviewM}[2]{\noindent\textbf{\hyperref[R#1M#2]{R#1M#2}}}

\def\BibTeX{{\rm B\kern-.05em{\sc i\kern-.025em b}\kern-.08em
    T\kern-.1667em\lower.7ex\hbox{E}\kern-.125emX}}
\begin{document}

\title{One Size Does NOT Fit All: On the Importance of Physical Representations for Datalog Evaluation}

\author{
\IEEEauthorblockN{Nick Rassau}
\IEEEauthorblockA{\textit{Johannes Gutenberg University} \\
Mainz, Germany \\
rassau@uni-mainz.de}
\and
\IEEEauthorblockN{Felix Schuhknecht}
\IEEEauthorblockA{\textit{Johannes Gutenberg University} \\
Mainz, Germany \\
schuhknecht@uni-mainz.de}
}
\maketitle

\begin{abstract}
Datalog is an increasingly popular recursive query language that is declarative by design, meaning its programs must be translated by an engine into the actual physical execution plan. When generating this plan, a central decision is how to physically represent all involved relations, an aspect in which existing Datalog engines are surprisingly restrictive and often resort to one-size-fits-all solutions. 
The reason for this is that the typical execution plan of a Datalog program not only performs a single type of operation against the physical representations, but a mixture of operations, such as insertions, lookups, and containment-checks. 
Further, the relevance of each operation type highly depends on the workload characteristics, which range from familiar properties such as the size, multiplicity, and arity of the individual relations to very specific Datalog properties, such as the ``interweaving'' of rules when relations occur multiple times, and in particular the recursiveness of the query which might generate new tuples on the fly during evaluation. 
This indicates that a variety of physical representations, each with its own strengths and weaknesses, is required to meet the specific needs of different workload situations. To evaluate this, we conduct an in-depth experimental study of the interplay between potentially suitable physical representations and seven dimensions of workload characteristics that vary across actual Datalog programs, revealing which properties actually matter. Based on these insights, we design an automatic selection mechanism that utilizes a set of decision trees to identify suitable physical representations for a given workload.

\end{abstract}

\vspace*{-0.2cm}
\section{Introduction}
\vspace*{-0.1cm}

Today, Datalog is being used in a wide array of applications, from program analysis~\cite{paper:souffle,paper:inca} to network monitoring~\cite{paper:networkMonitoring}, distributed computing~\cite{paper:dedalus,paper:cloudComputing}, and distributed storage~\cite{website:datomic}. What makes Datalog so popular is that it is declarative: the query describes only what should be computed, not how. From this, the Datalog engine generates a physical plan that is then executed on the dataset.   
When developing this plan, a central decision is how to represent all involved relations physically. This \textit{physical representation} comprises multiple aspects. 
First, there is the \textit{access type} that determines how tuples are stored and located. For instance, all tuples could be stored in flat tuple stores, which are then scanned to retrieve qualifying entries. Or, a combination of tuple stores and unclustered indexes could be created, used to locate and retrieve full tuples from the stores when needed. Alternatively, covered indexes could directly materialize all tuples internally, without the need for a separate tuple store altogether. 
On top, there is the choice of \textit{data structure}, which is particularly relevant for the indexes. Here, the usage of various ordered and unordered index structures, such as BTrees or hash tables, with highly different performance and space characteristics, is possible.

Despite having such a large amount of options, Datalog engines often implement rather static~\cite{paper:bddbddb} or rather limited choices~\cite{paper:recstep,lit:souffle:data_structures, paper:souffle:trie, paper:souffle:btree}. As an example, the state-of-the-art Datalog engine Soufflé~\cite{paper:souffle,paper:souffle:programrepair,paper:souffle:anotherpapercomparingtosouffle} represents each relation as a covered BTree. While this reduces the implementation effort and simplifies the code generation, it is (a)~unclear whether this decision is the best possible (or at least a good one) in all situations. Further, (b)~this design decision comes with side effects: If a relation occurs in multiple joins on different join keys, multiple BTrees must be created containing the tuples redundantly in different sort orders, which heavily increases the memory footprint. To work around this problem, Soufflé switches to unclustered BTrees for wide relations and stores the tuples only once separately. While applying this heuristic reduces the memory footprint for wide relations, it, of course, creates heavy random access, which might drastically deteriorate the query performance if the join selectivity is low.

\vspace*{-0.2cm}
\subsection{One-Size-Fits-All or Hand-Tailored?}
\vspace*{-0.1cm}

Still, there are relatable reasons for Datalog engines to be so simplistic in this regard.
First, the typical execution plan of a Datalog program not only performs a single type of operation against the physical representations, but a mixture of operations, such as insertions when new tuples are found, lookups for join processing, and containment checks to ensure set semantics. Consequently, one cannot simply go for the physical representation that performs well regarding \textit{one} specific operation, but has to consider its performance across a whole set of operations.  
Second, the relevance of each operation type highly depends on the workload characteristics, which range from familiar properties such as the size, multiplicity, and arity of the individual relations to very specific Datalog properties, such as the ``interweaving'' of rules when relations occur in multiple rules, and in particular the recursiveness of the query which might generate new tuples during evaluation. 
For instance, a workload that identifies high volumes of new tuples during its evaluation requires on the recursive relation a physical representation that handles individual append operations efficiently. 

These observations indicate that one-size-fits-all solutions are unlikely to be sufficient for Datalog evaluation. Instead, selecting from a catalog of physical representations with individual strengths and weaknesses depending on the faced workload characteristics is likely required. To evaluate this, we conduct an in-depth study of the interplay between different potentially suitable physical representations and the workload characteristics that vary across actual Datalog programs. This reveals which properties of physical representations and workloads actually matter. Based on these insights, we materialize a set of decision trees that could be the foundation of a physical query planner for future systems that incorporate a catalog of physical representations.
Our contributions are as follows:

\textbf{(1)}~To be able to study the relationship between different physical representations and workload characteristics, we develop a highly versatile Datalog engine called~\engine{} (\url{https://gitlab.rlp.net/rassau/playlog}). It follows an execution model inspired by Soufflé, while remaining highly flexible by being designed from the ground up to support a variety of physical representations. This flexibility does not incur any performance penalty, as \engine{} performs query compilation, yielding a result that is always a specialized C++ program that looks and runs like a handwritten plan.

\textbf{(2)}~In \engine{}, we integrated the following physical representations, where we consider a combination of two parts: (i)~The access type to the tuples of a relation, where we support to create a stand-alone covered index, an unclustered key-based index on top of a tuple store, an unclustered pointer-based index on top of a tuple store, and a stand-alone tuple store that is searched via a full-scan. (ii)~The used data structure to represent the index, where we support a naive sorted array, a B+-Tree, a radix tree, and a hash table. The tuple store is represented as a flat row-store. These structures have been used in the Datalog context before, as e.g. by Soufflé, RecStep~\cite{paper:recstep}, BigDatalog~\cite{paper:bigdatalog} and Myria~\cite{paper:myria}.  

\textbf{(3)}~We perform an in-depth study of the interplay between physical representations and workload characteristics. As workload characteristics, we consider seven dimensions that frequently vary in Datalog programs: (i)~The relation size, (ii)~the relation multiplicity, (iii)~the relation width, (iv)~the key width, (v)~the proportion of initialization and query effort, (vi)~the interweaving between and among rules in terms of reoccurring relations, and
(vii)~the recursiveness of a rule in terms of number of generated, unique, and new facts. We study these aspects on controllable synthetic workloads and four real-world workloads. 

\textbf{(4)}~We introduce a selection mechanism, which, based on signatures that capture relevant  properties of a workload, instantiates the physical representations of a given Datalog program based on a set of decision trees. To evaluate the effectiveness of the selection mechanism, we apply it to four real-world workloads and show that it is able to identify nearly ideal configurations that outperform a set of baseline systems.

Note that due to the high complexity of the topic, we focus in this study on single-node main-memory evaluation. Analyzing and optimizing distributed Datalog engines is beyond the scope of this paper, but are interesting topics that we plan to tackle in future work.

\vspace*{-0.2cm}
\section{Background and Motivation}
\label{sec:background}
\vspace*{-0.1cm}

We start by discussing how Datalog programs are evaluated. Based on that, we will discuss why the choice of physical representation is challenging yet important for Datalog evaluation.

\vspace*{-0.2cm}
\subsection{Datalog in a Nutshell}
\label{ssec:starting_point}
\vspace*{-0.1cm}

To get a sense of Datalog, we compute the reachability in a bidirectional graph. 
The starting node and the edges form the so-called input \textit{facts} and are materialized as tuples in corresponding relations such as \smalltt{Edges} = \smalltt{\{(0,1),(1,2),(1,4),(2,3),(4,5)\}} and \smalltt{Reachable} = \smalltt{\{(2)\}}, resembling the following graph:
\begin{center}
\includegraphics[width=0.28\textwidth,page=1]{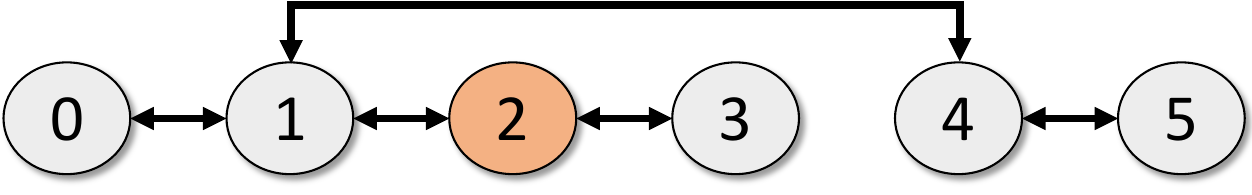}
\end{center}
\noindent To compute the reachability from node \smalltt{(2)}, we write a Datalog query consisting of two \textit{rules}, where the rules are logically evaluated one by one: 
\begin{footnotesize}
\begin{Verbatim}[commandchars=+\[\]]
Reachable(y) :- Reachable(+underline[x]), Edges(+underline[x],y)
Reachable(y) :- Reachable(+underline[x]), Edges(y,+underline[x])
\end{Verbatim}
\end{footnotesize}
Let us look at the evaluation of the first rule: The right side of the rule (aka its \textit{body}) joins the relation \smalltt{Reachable} with the relation \smalltt{Edges} on the first attribute by introducing a shared variable~\smalltt{x}. By introducing a second variable~\smalltt{y}, we capture the value of the second attribute of \smalltt{Edges} for each tuple that joined with a tuple from \smalltt{Reachable}. As this is what we are interested in in our reachability computation, we insert the resulting values~\smalltt{y} as potentially new tuples into the \smalltt{Reachable} relation on the left side of the rule (aka its \textit{head}). 
As our head relation is also part of the body, this results in a \textit{recursive evaluation}, in which the rule is repeatedly evaluated until no new facts are generated. 
Note that \smalltt{Edges} is also called an extensional database (EDB) as it contains facts to start with and remains static, while \smalltt{Reachable} is called an intensional database (IDB), as it potentially receives new facts.

\vspace*{-0.2cm}
\subsection{Program Evaluation}
\vspace*{-0.1cm}

When evaluating Datalog programs, we distinguish between non-recursive and recursive rules. 
Evaluating a non-recursive rule of the form \smalltt{D(i,l) :- A(i,j), B(j,k), C(k,l)} is carried out by: First, initializing all relations (creating the physical representations and potentially filling them). Second, evaluating the body as a sequence of joins between the body's relations based on the join keys, which potentially produces facts; third, inserting these facts into the head relation if they are new. 
Evaluating a recursive rule such as \smalltt{A(i,l) :- A(i,j), B(j,k), C(k,l)} involves more steps, where the de-facto standard technique~\cite{paper:survey} is \textit{semi-naive bottom up evaluation}, implemented for instance by Soufflé, BigDatalog, SociaLite~\cite{paper:socialite}, or RecStep~\cite{paper:recstep}. The core idea is to translate the recursive rule into a sequence of non-recursive evaluations that terminate when no new facts are produced anymore. To keep overhead low, each iteration of the sequence bases its evaluation only on the facts discovered in the previous iteration. 
Algorithm~\ref{alg:physical_execution_in_souffle} shows in pseudo-code how the program is evaluated, where the joins are carried out as a sequence of (index)-nested-loop-joins as done by Soufflé: 
For the IDB~\Abase{}, a relation called \Adelta{} is created additionally. At all times, this relation contains only the facts that were newly generated in the previous iteration, excluding potential duplicates that already exist in \Abase{}.
To kick off the evaluation, \Adelta{} is filled with all initial tuples of \Abase{}. Then the main loop starts the first iteration, in which it evaluates the body. Each fact found is, however, not directly inserted into \Abase{}, but into a third relation called \Anew{}, which is freshly created in each iteration.   
After all facts of this iteration have been produced, the \Adelta{} relation for the next iteration must be constructed, which receives all tuples of \Anew{}. Before going into the next iteration, \Adelta{} is  merged into \Abase{}. 
If \Adelta{} remained empty, \Abase{} is returned as the final result.

\begin{algorithm}[h!]
\caption{Semi-naive bottom up evaluation. Colors mark the different performance-critical operations.}
\label{alg:physical_execution_in_souffle}
\tcp{A(i,l) :- A(i,j), B(j,k), C(k,l)}
\label{line:creation}\textcolor{BrickRed}{\AbaseInCode{}.create(on key~0\_1)}, \textcolor{RoyalBlue}{\AbaseInCode{}.bulk\_load(inputA)}\\
\textcolor{BrickRed}{B.create(on key~0)}, \textcolor{RoyalBlue}{B.bulk\_load(inputB)}\\
\textcolor{BrickRed}{C.create(on key~0)}, \textcolor{RoyalBlue}{C.bulk\_load(inputC)}\label{line:creation_end}\\
\textcolor{BrickRed}{\AdeltaInCode{}.create(on key~0\_1)}, 
\textcolor{RoyalBlue}{\AdeltaInCode{}.bulk\_load(inputA)} \\
\While{\textnormal{!\AdeltaInCode{}.empty()}}{
  \textcolor{BrickRed}{\AnewInCode{}.create(on key~0\_1)}\\
  \For{\textnormal{\textcolor{red}{A\_entry : \AdeltaInCode{}}}} { 
    \textcolor{BlueGreen}{B\_entries = B.probe((A\_entry[1]))} \label{line:probe1} \\
    \For{\textnormal{\textcolor{red}{B\_entry : B\_entries}}} {  \label{line:probe_result_iteration}
        \textcolor{BlueGreen}{C\_entries = C.probe((B\_entry[1]))} \label{line:probe2} \\
        \For{\textnormal{\textcolor{red}{C\_entry : C\_entries}}} {
            fact\_found = (A\_entry[0], C\_entry[1]) \\
            \If{\textnormal{!\textcolor{Orange}{\AbaseInCode{}.contains(fact\_found)}}} { \label{line:dedup_start}
                \If{\textnormal{!\textcolor{Orange}{\AnewInCode{}.contains(fact\_found)}}} {
                    \textcolor{Orchid}{\AnewInCode{}.append(fact\_found)} \\
                }
            }
        }
    }
  }
  \textcolor{ForestGreen}{\AnewInCode{}.finished\_append()}\\
  \For{\textnormal{\textcolor{red}{\AnewInCode{}\_entry : \AnewInCode{}\_A}}} {
    \textcolor{Orchid}{\AbaseInCode{}.append(\AnewInCode{}\_entry)} \\
  }
  \textcolor{ForestGreen}{A.finished\_append()}\\
  \AdeltaInCode{}.move\_from(\AnewInCode{}) \label{line:dedup_end} \\
}
\Return A
\end{algorithm}

\vspace*{-0.3cm}
\subsection{Performance-critical Operations are Everywhere!}
\label{ssec:physical_execution_souffle}
\vspace*{-0.1cm}

As shown in Algorithm~\ref{alg:physical_execution_in_souffle}, performance-critical operations on the physical representations, which we highlighted in different colors for different operations, are omnipresent: 

Initially, \Abase{}, \Adelta{}, \Anew{}, \smalltt{B}, and \smalltt{C} must be created from scratch, where all except \Anew{} are then bulk-loaded with a potentially large number of input tuples. As these input tuples typically arrive on the fly~\cite{paper:souffle,paper:difficultCompiletimeOptimization,paper:difficultCompiletimeOptimization2}, this  becomes part of the actual program evaluation. Consequently, if the construction is the dominant part of the program evaluation, a physical representation that is very efficient in this regard (e.g., by not having to build an auxiliary structure or by exploiting access locality during bulk-loading) could be the better choice even though it may lack performance over its competitors during later usage.   
In the sequence of joins, we have to differentiate between the probing part into the physical representation of the right side of each join pair as well as the subsequent iteration over the probe result that is required to go on in the join sequence.
Depending on the number of tuples, the number of matching join partners, and the number of duplicates per match, the effort can vary widely across parts, which consequently favors physical representations optimized for the dominant operation. 
After identifying a found fact, deduplication must be performed by performing contains checks on both \Abase{} and \Anew{}, requiring the physical representations to test for the containment of a tuple efficiently. 
After identifying that a fact is genuinely new, it must be inserted into the \Anew{} relation. Consequently, if a large number of new facts are found, a physical representation with support for fast individual append operations can likely handle this case better than an inflexible one. 
Finally, the content of \Anew{} must be copied into the base relation and the current \Anew{} relation becomes the next \Adelta{} relation. The former is another form of bulk-loading, although it now happens into an already populated physical representation and may modify it in a scattered fashion. The latter can be carried out as a cost-free ``rebranding'' if the physical representation of \Anew{} matches that of \Adelta{}, which we assume in the following. 
In addition to how operations are balanced during evaluation, properties such as schema widths, key widths, and the potential for sharing physical representations likely also influence the choice of physical representation. In Section~\ref{sec:non_recursive_evaluation}, we will study these relationships in depth.

\begin{table*}[h!]
    \centering
    \scriptsize
    \begin{tabular}{C{0.9cm}cccc}\toprule
             & \CI{} & \UKI{} & \UPI{} & \FS{} \\
             \midrule
    \ARRAY{}\textbf{(++)} & T[] & \textless K,T*\textgreater[] & T*[] & -\\
       \BP{} & tlx::btree\_multiset\textless T\textgreater & tlx::btree\_multimap\textless K,T*\textgreater & tlx::btree\_multiset\textless T*\textgreater & -\\
       \HT{} & google:dense\_hash\_map\textless K,std::vector\textless T{\textgreater}\textgreater & google:dense\_hash\_map\textless K,std::vector\textless T*{\textgreater}\textgreater & google:dense\_hash\_map\textless T*,std::vector\textless T*{\textgreater}\textgreater & -\\
       \RX{} & seq::radix\_map\textless T,std::vector\textless T{\textgreater}\textgreater & seq::radix\_map\textless K,std::vector\textless T*{\textgreater}\textgreater & seq::radix\_map\textless T*,std::vector\textless T*{\textgreater}\textgreater & -\\
       \BL{} & - & - & - & T[][] \\\bottomrule
    \end{tabular}
    \caption{All supported combinations of access types and data structures along with their internal representation. \smalltt{T} is the type of the tuple of the relation, whereas \smalltt{K} is the type of the key. For all \UKI{} and \UPI{} combinations, \smalltt{T*} points into a separate \BL{}.}
    \label{tab:physical_representations}
    \vspace*{-0.4cm}
\end{table*}

\vspace*{-0.2cm}
\section{Physical Representations}
\label{sec:physical_representations}
\vspace*{-0.1cm}

As physical representations, we consider four access types and five data structures, yielding a total of 13 combinations.

\textbf{Access types}. The access type determines where tuples are stored, how they are indexed (if at all), and how they are retrieved. 
The first access type is the covered index (\CI{}), which materializes all tuples directly in the underlying data structure, so no separate store is required to host them. The downside of \CI{} shows up if the respective relation participates in different joins on different join keys where one join key is not a prefix of the other. This leads to the creation of multiple \CI{}s, which redundantly store the tuples of the relation.
To counter this, one can fall back to unclustered pointer-based indexes (\UPI{}s) for wide relations with many different join keys. Therein, all tuples are stored in a separate row-store and are only referenced by the underlying index structure. This index structure does not materialize explicit keys and map them, but only stores tuple addresses (hence the name pointer-based), which must be dereferenced at probe time to reconstruct the keys. While this results in expensive random accesses, the reason for such a design is that wide compound keys are rather common in Datalog programs~\cite{paper:souffle, paper:datalogBigRulesOccure, paper:Vadalog}. 
Still, we also support unclustered key-based indexes (\UKI{}s), which explicitly materialize keys to avoid random accesses in the tuple row-store at probe time and may be a good choice for narrow keys. 
As a fourth access type, we support a full scan (\FS{}) that answers a probe by scanning and filtering for qualifying entries.

\textbf{Data structures}. As data structures, we support the following options: a B+-Tree (\BP{}), which stores all tuples at the leaf level and supports efficient range scans. We use the popular implementation from the TLX library~\cite{online:tlx} in its default configuration. 
Further, we support a radix tree (\RX{}) in our evaluation in the form of \smalltt{seq::radix\_tree} from the seq library~\cite{online:seq}. 
This variant implements a variable arity radix tree (VART), which adaptively uses different node sizes depending on the sparsity of the dataset, similar to ART~\cite{lit:art} or JudyArray~\cite{lit:judyarray}. 
To include an ordered data structure with a minimal memory footprint, we consider a sorted, consecutive array of tuples (\ARRAY{}). To handle insertions efficiently, which is challenging for dense arrays, we integrate two optimizations: First, we use the system call \smalltt{mremap()} to increase the array on demand at the granularity of $2$MB~chunks (the size of a huge page). This avoids copying physical memory and relocates virtual memory only if necessary. Second, when \smalltt{append()} operations are happening, we do not sort in each tuple directly. Instead, we first store them in an unsorted area at the end of the array and then sort them in after the sequence of \smalltt{append()} operations has finished. Note that this can slow down \smalltt{contains()} checks interleaved with \smalltt{append()} operations, as they must scan the unsorted area to answer the request if they cannot locate the tuple in the sorted area.
To counter this problem, we include an alternative: turning off the \smalltt{contains()} semantic while \smalltt{append()} operations are happening. As a result, duplicates may be introduced into the data structures, which are then removed once the evaluation indicates that the sequence of appends has completed. This removes the quadratic complexity from interleaved \smalltt{contains()} and \smalltt{append()} operations, but requires an additional post-append pass via \smalltt{std::unique}. We differentiate between the two variants in the following as \ARRAY{} (normal \smalltt{contains()} behavior) and \ARRAYALTERNATIVE{} (deactivated \smalltt{contains()} during \smalltt{append()} sequences), where both behave the same under probes. 
Since a central operation in rule evaluation is performing equi-joins, we further include a hash table (\HT{}) as a natural candidate for efficient point-lookups. We use \smalltt{google::dense\_hash\_map()} from Google's sparsehash library~\cite{online:sparsehash}, which uses open addressing with linear probing internally and maintains a maximum default load factor of~$0.5$. As hash function, we use the 64-bit variant of MurmurHash from the SMHasher suite~\cite{online:murmur}. Due to a lack of sortedness, the \smalltt{probe()} operation of \HT{} does not support range-lookups, so a single physical representation can not be used in multiple joins with prefix join keys. As the underlying structures of both \HT{} and \RX{} do not support duplicate keys, we use a vector of tuples as a value to still handle this case. 
Further, we also include a flat data structure, a row store (\BL{}). As certain access types might reference tuples in this row store, we do \textit{not} organize it as a dense array resized with \smalltt{mremap()}, as in \ARRAY{}, since this might move the virtual memory area and break references. Instead, we organize \BL{} as a resizing list of consecutive memory chunks, each of size $2$MB.   

Note that all our data structures support storing duplicates by default and rely on the execution to enforce set semantics on all relations via appropriate \smalltt{contains()} checks. This grants us flexibility in testing different strategies for building up the recursive relation, as we will discuss in Section~\ref{ssec:deduplication}. Also, to measure the exact memory footprint of \BP{}, \RX{}, and \HT{}, we equip the internal indexes with a custom allocator that tracks all allocations. 
Table~\ref{tab:physical_representations} shows all $13$~meaningful combinations of access types and data structures.

\vspace*{-0.2cm}
\section{Interplay of physical representations and workload characteristics}
\label{sec:non_recursive_evaluation}
\vspace*{-0.1cm}

In the following, we will analyze the interplay of physical representations and seven different workload characteristics. To handle the complexity of Datalog evaluation, we break down the most important aspects of Datalog into individual experiments. In Section~\ref{sec:automatic_selection}, we will then materialize these individual findings on a set of decision trees on which our selection mechanism operates.

We perform all upcoming experiments on an Intel i9-12900K running at up to 5.2GHz. The CPU has 30MB of shared L3 cache, whereas a single performance core has 1.25MB of L2 cache and 32KB of L1 cache available. The machine is equipped with 128GB of DDR4 main memory and runs a 64-bit Arch Linux with kernel 6.13.4. Transparent huge pages are enabled and may be used for memory allocation. We compile \engine{} using g++ in version~14.2.1 on level~O3. All reported runtimes are the average of three runs. Further, all relations used in the following evaluation store $8$B~unsigned integers for all of their attributes. Note that Soufflé also supports only on fixed-size data types in their core engine; variable-sized data types like strings can still be processed by encoding/decoding the values to/from integers via a separate dictionary.

\vspace*{-0.2cm}
\subsection{\mbox{Scaling Behavior of the Body Evaluation}}
\label{ssec:join_performance}
\vspace*{-0.1cm}

To get a first impression of the different physical representations, we start with an analysis of their scaling behavior when varying the number of contained and hence probed tuples. For now, we assume the relations have been bulk-loaded already and focus solely on the body evaluation, where we evaluate a body of the form \smalltt{R(\underline{x},y), S(\underline{x},z)}, which produces a checksum from the \smalltt{y} and \smalltt{z} attributes of the join results.
We fix the outer relation~\smalltt{R} to $1.25$MiB and vary the probed-into inner relation~\smalltt{S} from $10$MiB over $100$MiB to $1{,}000$MiB (memory footprint of the raw tuples), where we fill both relations with densely selected but shuffled tuples. While varying~\smalltt{S}, we make sure that each tuple of \smalltt{R} finds exactly two join partners in~\smalltt{S} to keep the number of join results and hence the actual join effort fixed. This allows us to evaluate how the different physical representations handle the scaling on the probe side.

\begin{figure}[h!]
    \centering 
     \begin{subfigure}{\columnwidth}
        \centering
         \includegraphics[width=\linewidth]{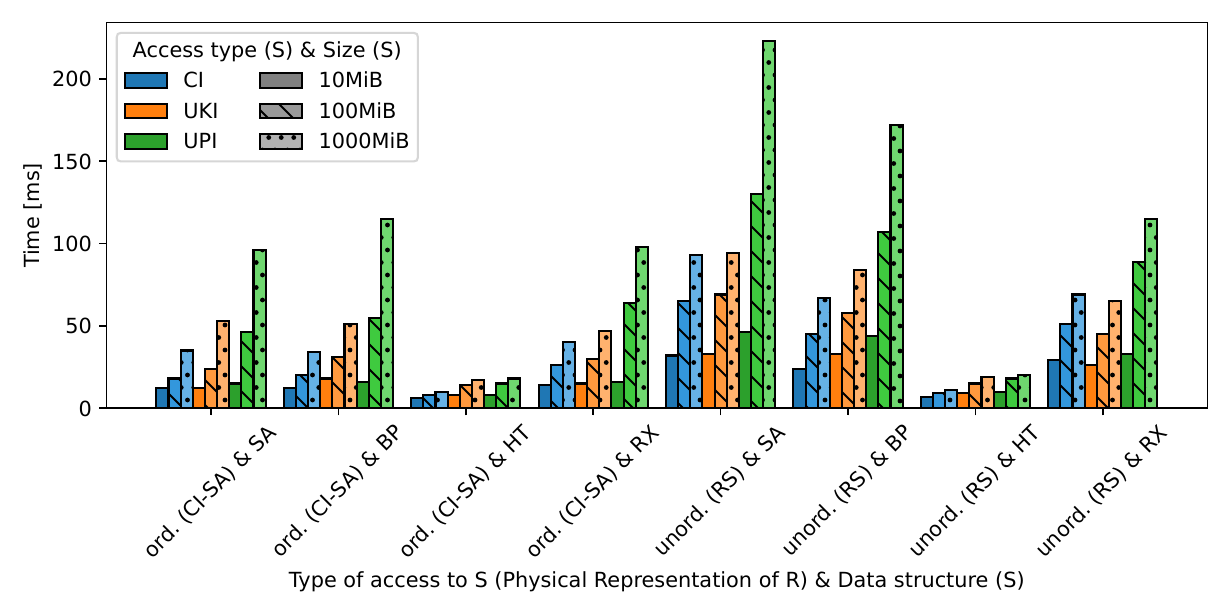}
         \caption{Body evaluation time.}
         \label{fig:distance_between_lookups:query}
     \end{subfigure}
     \begin{subfigure}{\columnwidth}
        \centering
         \includegraphics[width=\linewidth]{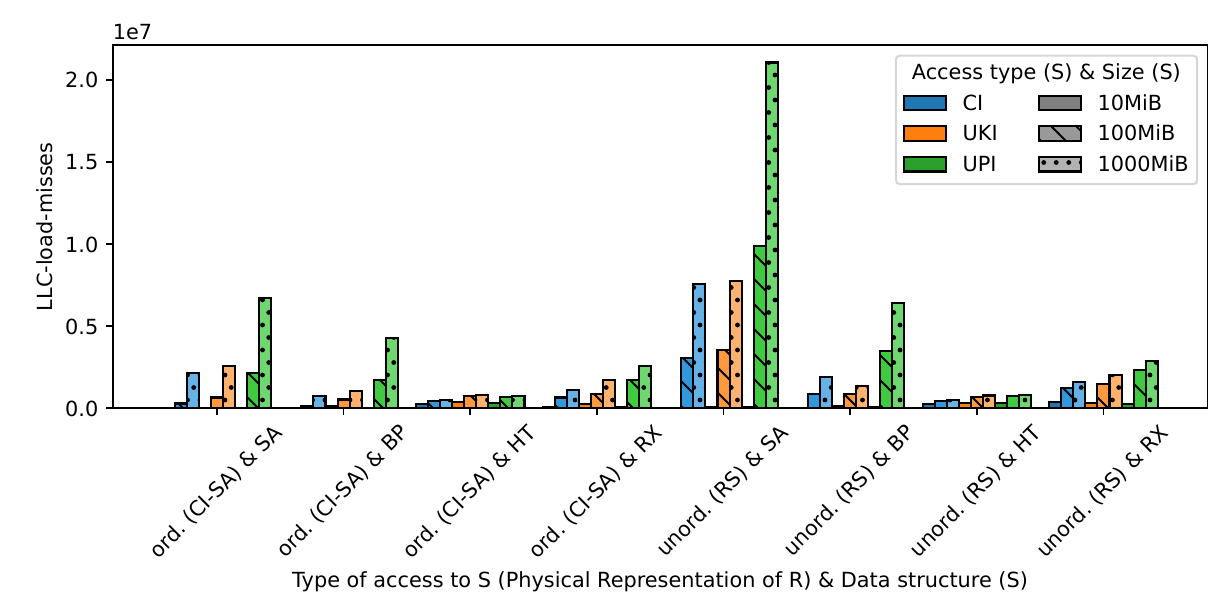}
         \caption{Body evaluation L3 cache misses (LLC-load-misses counter).}
         \label{fig:distance_between_lookups:cache_misses}
     \end{subfigure}   
     \begin{subfigure}{\columnwidth}
        \centering
         \includegraphics[width=\linewidth]{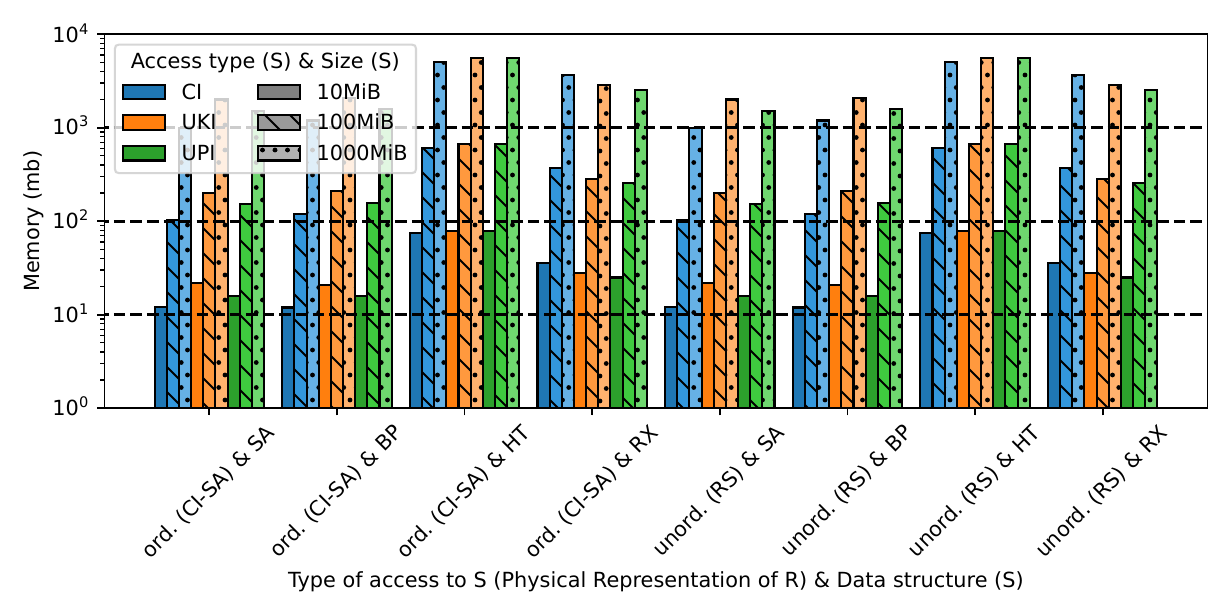}
         \caption{Memory footprint.}
         \label{fig:distance_between_lookups:memory}
     \end{subfigure}
     \vspace*{-0.5cm}
        \caption{Scaling behavior of different physical representations.}
        \label{fig:distance_between_lookups}
        \vspace*{-0.2cm}
\end{figure}

Figure~\ref{fig:distance_between_lookups:query} shows the total body evaluation time for the different setups. For the inner relation~\smalltt{S}, we present all meaningful data structure and access type combinations, whereas for the outer relation~\smalltt{R}, we show only two representatives to maintain a readable plot. These are \CI{}-\ARRAY{}, which generates ordered probes into~\smalltt{S}, and $\BL{}$, which can be considered the equivalent for generating unordered probes. 
From the results, we can observe significant performance differences between the tested physical representations. Focusing on the access type first, we can see that \CI{} generally outperforms \UPI{}, whereas \UKI{} positions itself in between. This ordering is reasonable, as \CI{} avoids random accesses to a separate store entirely, while \UKI{} performs them into the underlying \BL{} to retrieve the tuples. \UPI{} suffers even more under random accesses, as it must also access the underlying \BL{} during the index traversal, i.e., to perform the key comparisons. The performance differences are also more pronounced the larger~\smalltt{S} becomes: While for $10$MiB, \CI{} and \UPI{} have a performance difference of less than $2\times$, for $1{,}000$MiB, it grows to up to $8\times$, indicating that \UKI{} and especially \UPI{} should be used with caution when representing larger relations. This effect occurs because the 30MB L3 cache and the 32 slot L1-TLB for huge pages (covering 64MB) that caches address translations quickly become too small relative to~\smalltt{S} to prevent expensive misses during random accesses. This is confirmed by Figure~\ref{fig:distance_between_lookups:cache_misses}, which plots the \smalltt{LLC-load-misses} counter obtained in a profiling run of the body evaluation using \smalltt{perf}, where we can clearly see the correlation to the runtime. The same holds for the dTLB-misses counter, which we do not show due to space constraints.

If we add data structures to the discussion, we can make several interesting observations: First, \BP{} is not the best-performing option. For ordered probes, it is the worst option and is even outperformed by \ARRAY{}, which exploits the high temporal locality of the access pattern but incurs less structural overhead and hence remains longer cache resident. \RX{} also performs slightly better as it is comparison-free and can index the dense tuples very effectively. As expected, \HT{} outperforms all other variants significantly and is exceptionally robust in combination with \UKI{} and \UPI{} as it requires less memory accesses per probe and hence causes less cache and TLB misses  -- this is good news, as using \UKI{} or \UPI{} is typically not a free choice, but forced by a space constraint. In such a case, it is pleasant that we are still able to achieve good performance if combined with \HT{}. On the downside, as we can see in Figure~\ref{fig:distance_between_lookups:memory} showing the memory footprint, \HT{} has the largest footprint of all structures, limiting its usability in this context again (at least when used with a maximum load factor of $0.5$~as done here).    
Under unordered probes, all comparison-based data structures suffer, with \BP{} now the second-worst option compared to \ARRAY{}. As expected, \HT{} remains essentially unaffected by the access pattern and is therefore the most reasonable choice in this situation.  
Overall, these initial findings suggest that relying solely on a single data structure, such as \BP{}, as a one-size-fits-all solution overlooks interesting opportunities. Also, the access type has a significant performance impact on ordered structures and must therefore be aligned with the data structure.

\vspace*{-0.2cm}
\subsection{Body Evaluation: Probe vs Probe-Result Iteration}
\label{ssec:probe_vs_probe_result_iteration}
\vspace*{-0.1cm}

After studying the join operation in the body evaluation from an end-to-end perspective, let us zoom in to examine the impact of the probing and probe-result iteration parts. For example, for \CI{}-\BP{}, the former part would be the tree traversal required to determine the lower bound of qualifying entries, where the latter part would be the scanning of leaf nodes until the upper bound is found.
To adjust the effort that goes into each phase, we now keep the size of the inner relation~\smalltt{S} fixed to $100$MiB and vary the number of contained duplicates -- the more duplicates, the less effort goes into the probe phase and the more into the probe-result iteration. We test the three cases where each key occurs only once, $10$~times, and $100$~times. In the outer relation~\smalltt{R}, each key always occurs once, such that the number of join partners is fixed for all three experiments.

\begin{figure}[h!]
    \centering
     \begin{subfigure}{0.32\columnwidth}
        \centering
         \includegraphics[width=\linewidth]{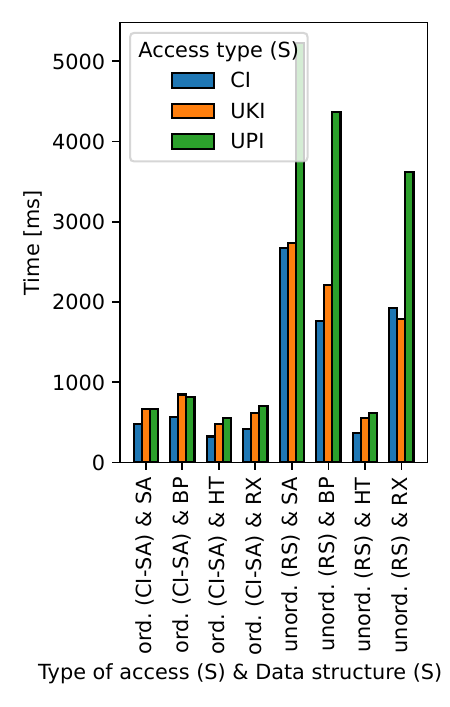}
         \caption{Each key 1$\times$.}
         \label{fig:varying_multiplicity:6553600}
     \end{subfigure}
     \begin{subfigure}{0.32\columnwidth}
        \centering
         \includegraphics[width=\linewidth]{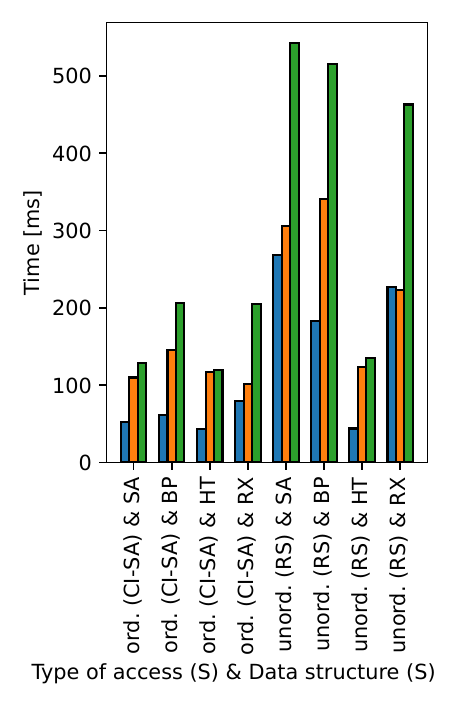}
         \caption{Each key 10$\times$.}
         \label{fig:varying_multiplicity:65536}
     \end{subfigure}
     \begin{subfigure}{0.32\columnwidth}
        \centering
         \includegraphics[width=\linewidth]{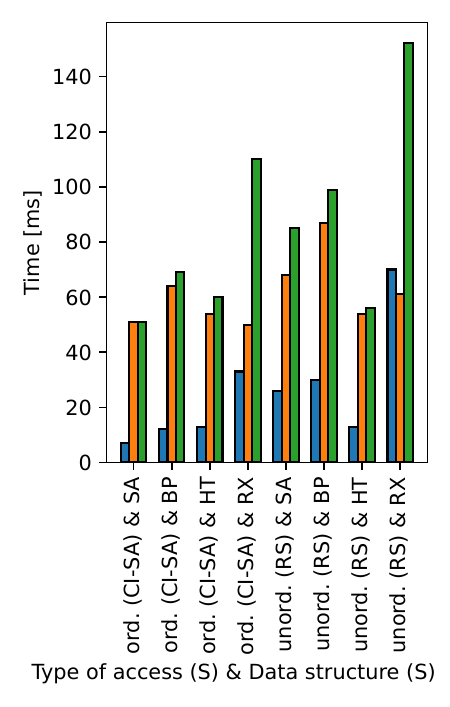}
         \caption{Each key 100$\times$.}
         \label{fig:varying_multiplicity:65536}
     \end{subfigure}
     \vspace*{-0.2cm}
        \caption{Splitting the join in probe and probe-result iteration while varying the number of duplicates in~\smalltt{S}.}
        \label{fig:varying_multiplicity}
        \vspace*{-0.3cm}
\end{figure}

\begin{figure*}[h!]
    \centering
    \begin{subfigure}{.32\linewidth}
    \includegraphics[width=.95\linewidth]{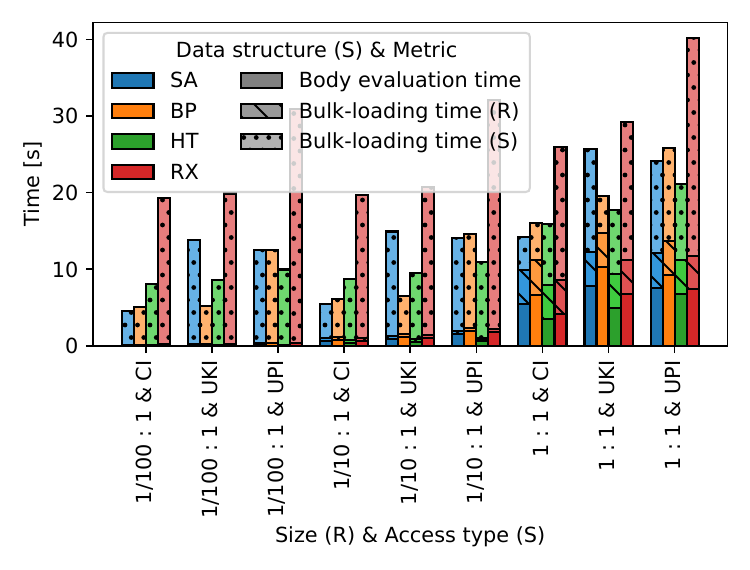}
    \caption{Physical representation (R): \CI{}-\ARRAY{}.}
    \label{fig:construction_vs_query_time:CI-SA}
    \end{subfigure}
    \begin{subfigure}{.32\linewidth}
    \includegraphics[width=.95\linewidth]{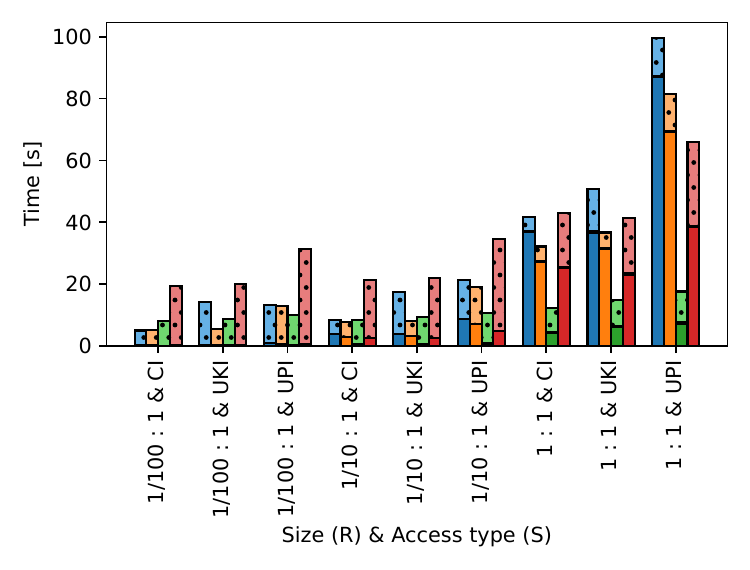}
    \caption{Physical representation (R): \BL{}.}
    \label{fig:construction_vs_query_time:RS}
    \end{subfigure}
    \begin{subfigure}{.32\linewidth}
    \includegraphics[width=.95\linewidth]{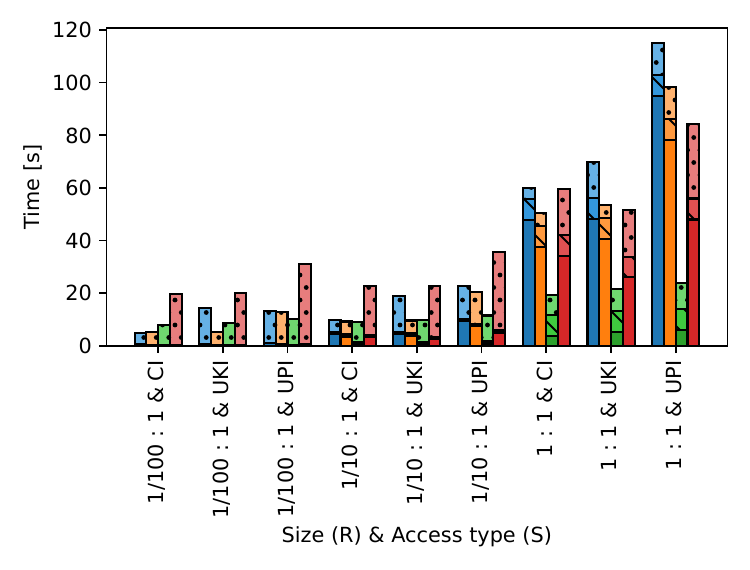}
    \caption{Physical representation (R): \CI{}-\HT{}.}
    \label{fig:construction_vs_query_time:CI-HT}
    \end{subfigure}
    \caption{Bulk-loading vs body evaluation.}
    \label{fig:construction_vs_query_time}
    \vspace*{-0.3cm}
\end{figure*}

Figure~\ref{fig:varying_multiplicity} shows the body evaluation time for all three configurations. We can see that the amount of duplicates not only impacts the runtime, where more duplicates significantly decrease the time, but also the relative performance of the individual physical representations to each other, showing that the representations handle the shift differently well. 
With an increase in duplicates, the advantage of ordered probes decreases as a larger portion of effort goes into the probe-result iteration. Hence, under lots of duplicates, an engine can rather afford to probe unordered into an ordered data structure like \ARRAY{}, \BP{}, or \RX{}, instead of enforcing the usage of the robust but more expensive to construct and space inefficient~\HT{}.
Regarding the data structure, the relation between all four variants changes with the number of duplicates, where \ARRAY{}, \HT{}, and \RX{}, which store all duplicates consecutively, benefit more from an increasing number of duplicates due to more effective prefetching than \BP{}, which scans individual leaf nodes.
When focusing on the access type, we can see that the relative difference between \CI{}, \UKI{}, and \UPI{} highly varies with the relation between probe and probe-result iteration effort. For a low number of duplicates resulting in a focus on the probe phase, the performance of \CI{} and \UKI{} is very similar while \UPI{} is heavily inferior (for unordered probes). For a high number of duplicates, \CI{} is significantly superior over both \UKI{} and \UPI{}, which now perform very similarly. This shows that in the former case, \UKI{} is a performance-neutral alternative to \CI{}, while in the latter case, it is clearly not due to a high number of cache and TLB misses during tuple lookups, and \CI{} should be used if affordable space-wise.

\vspace*{-0.2cm}
\subsection{Bulk-loading Physical Representations vs Body Evaluation}
\label{ssec:init_vs_body_eval}
\vspace*{-0.1cm}

So far, we looked solely at the body evaluation. However, in the Datalog world, EDBs often arrive on-the-fly, and hence, the bulk-loading of physical representations becomes a part of the whole program evaluation. 
Depending on the workload characteristics, bulk-loading or body evaluation might be dominant over the other -- the question in this regard is whether the choice of physical representation should depend on this relation.
To find out, we start with the setup from the previous subsections and change the following: Instead of varying the size of the inner relation~\smalltt{S}, we now keep~\smalltt{S} fixed to $1{,}000$MiB and vary the size of the outer relation~\smalltt{R} from $10$MiB ($\frac{1}{100}$:1 size relationship between \smalltt{R} and \smalltt{S}) over $100$MiB ($\frac{1}{10}$:1) to $1{,}000$MiB (1:1). By this, we fix the bulk-loading effort of the relation we are probing into while varying the body evaluation effort connected to it. This allows us to adjust whether one phase dominates over the other.

Figure~\ref{fig:construction_vs_query_time} shows the results for three different physical representations of \smalltt{R}, where we test \CI{}-\ARRAY{} as the most lightweight method in terms of bulk-loading effort for generating ordered probes, and both \BL{} and \CI{}-\HT{} as lightweight respectively heavyweight representatives for generating unordered probes. We split the total time into the bulk-loading time of \smalltt{R} and \smalltt{S} and the body evaluation time. 
Let us first focus on the case of a 1:1 size relationship between~\smalltt{R}~and~\smalltt{S}. Even in this case, where the body evaluation effort is high, the bulk-loading time makes a significant portion of the total time, showing that bulk-loading cannot generally be ignored, but must be considered if the querying effort and initialization effort is somewhat balanced. 
Even more, we can see that for ordered probes (\CI{}-\ARRAY{}), the bulk-loading time dominates the total time for all physical representations of~\smalltt{S}, while for unordered probes (\BL{} and \CI{}-\HT{}), this is not yet the case at a 1:1 relation. This shows that the probe order must also be factored in when deciding on a specific physical representation.
If we inspect the bulk-loading time of~\smalltt{S}, we can see that it highly varies for the different physical representations: In terms of the access type, the initialization time for \UPI{} in combination with ordered data structures is significantly higher than for \CI{} and \UKI{}, which require around the same time. This is due to the frequent random accesses into the underlying row-store during sorting and insertion at the right place, while for \HT{}, only one trip to the store during the hash computation is required. Focusing on the data structure, \ARRAY{} has naturally the shortest initialization time as it only sorts, followed by \BP{}, \HT{}, and \RX{}. \HT{} is surprisingly expensive although it does not involve a sorting step; however, it suffers from random accesses during bulk-loading. When also taking the $\frac{1}{100}$:1 and $\frac{1}{10}$:1 relationships into account, we can observe the inverse relationship between the body evaluation and bulk-loading time, showing that indeed different decisions should be made depending on whether one or the other dominates. We see this specifically for the more expensive unordered probes (\BL{} and \CI{}-\HT{}), where for a relationship of $\frac{1}{100}$:1, \CI{}-\ARRAY{} is the best choice due to its low initialization time. In contrast, for a relationship of 1:1, it becomes the worst option due to its high body evaluation time.
Also note that not using an index for \smalltt{R} but only \BL{} can significantly boost the evaluation due to a faster construction.

\vspace*{-0.2cm}
\subsection{Schema Width and Key Width}
\label{ssec:schema_key_width_rule_length}
\vspace*{-0.1cm}

The previously studied body \smalltt{R(\underline{x},y), S(\underline{x},z)} performs a join on \textit{one} attribute between \textit{two} relations that are \textit{two} attributes wide each. Since both schema and key width are also parameters that can highly vary, we now adjust them independently and study their effect on the physical representations.

\begin{figure}[h!]
    \centering
     \begin{subfigure}{\columnwidth}
        \centering
         \includegraphics[width=.9\linewidth]{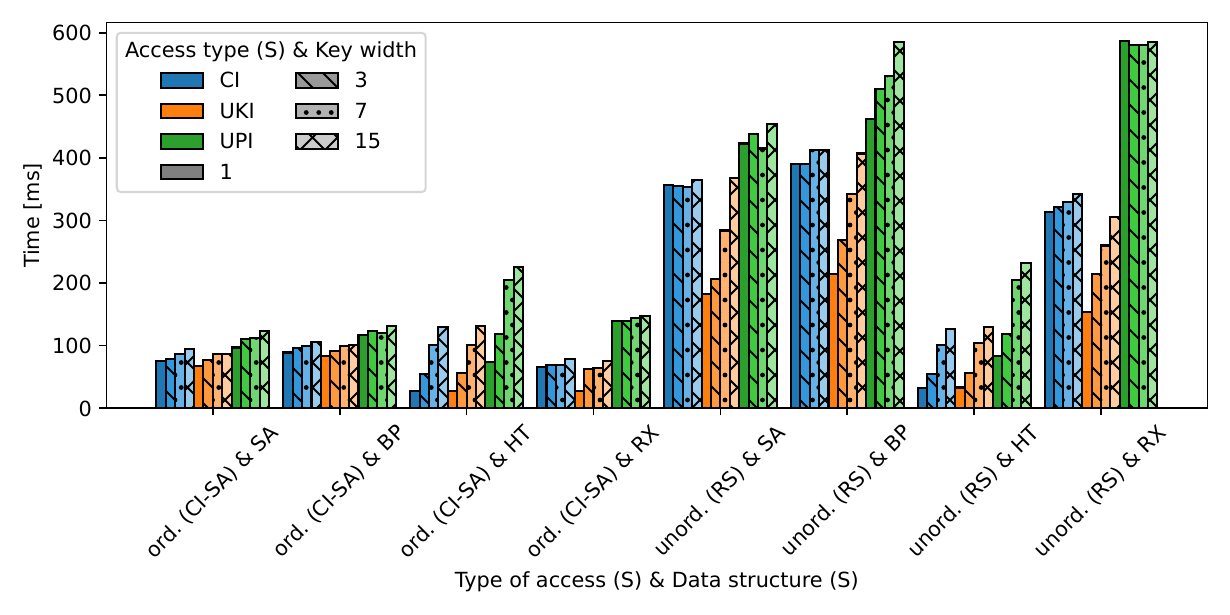}
         \caption{Impact of the key width.}
         \label{fig:width:key}
     \end{subfigure}
     \begin{subfigure}{\columnwidth}
        \centering
         \includegraphics[width=.9\linewidth]{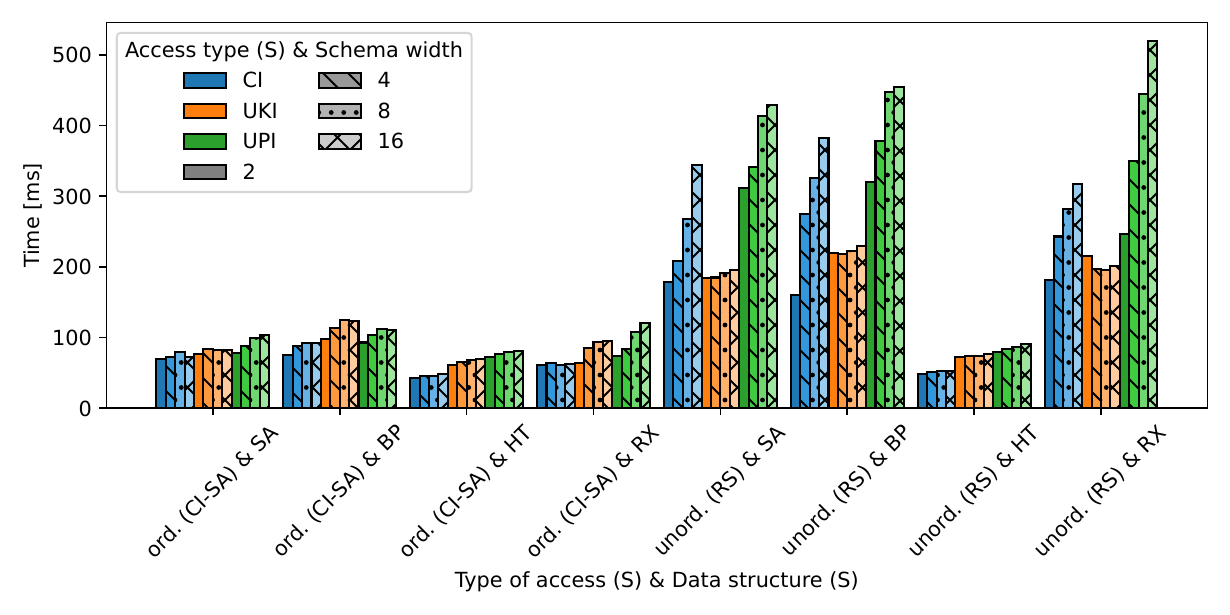}
         \caption{Impact of the schema width.}
         \label{fig:width:schema}
     \end{subfigure}
        \caption{Impact of the key width and the schema width.}
        \label{fig:width}
        \vspace*{-0.5cm}
\end{figure}

We start with the key width, where we extend the schema of both \smalltt{R} and \smalltt{S} to $16$~attributes and then vary the width of the join key among $1$, $3$, $7$, and $15$. We populate \smalltt{R} and \smalltt{S} with $1$M tuples each ($122$MiB per relation). For each configuration, we build the index on \smalltt{S} accordingly, i.e., on a (compound) key of the corresponding width. 
Figure~\ref{fig:width:key} shows the body evaluation time, again under ordered and unordered probes. 
We can see that for all variants, the runtime increases with an increase in key width, however, at a different intensity. \ARRAY{} and \BP{} in combination with \CI{} and \UPI{} are largely unaffected by the key width, as they do not have to explicitly materialize the keys internally (as they store the full tuples, requiring only $124$MiB), but only suffer from the additional comparisons. This changes when being combined with \UKI{}, where the performance under unordered probes correlates with the key width. However, this correlation can also be an advantage, if the key width is significantly smaller than the schema width: For \ARRAY{} and \BP{}, \UKI{} for a key width of~$1$ is around twice as fast as \CI{}, showing that an unclustered data structure can also have performance benefits. For \RX{}, independent from the access type, a wider key results in a longer traversal path in the radix tree, and hence, the sensitivity for the key width is high here. The same holds for \HT{} in combination with \CI{} and \UKI{}, where the key must be explicitly materialized for each slot of the hash table.
This renders \UKI{} from a small-sized index with $140$MiB for a key width of $1$ to $248$MiB for a key width of $15$. For \HT{} more space is required: $190$MiB with key width~$1$ for all access types up to $415$MiB for \CI{} and \UKI{} for key width~$15$. As expected, the key width has no impact on the memory consumption while using \UPI{}.
In Figure~\ref{fig:width:schema}, we now vary the schema width of the inner relation~\smalltt{S} from $2$ attributes to $16$~attributes, while keeping the schema of~\smalltt{R} fixed to $2$~attributes where the join is performed on the first attribute. 
We can observe that for ordered probes, all variants remain relatively unaffected by an increase in schema width. For unordered probes, it is different: While \UKI{} remains very stable (as it handles only keys), the \CI{} and \UPI{} variants deteriorate in performance when combined with the ordered data structures \ARRAY{}, \BP{}, and \RX{}, as they experience random accesses on a significantly larger memory region, causing more cache and TLB misses. Again, we observe the benefit of \UKI{}: Because it is independent of the schema width, the body evaluation time remains stable as the schema width increases, significantly outperforming \CI{}. 
For the memory that is accessed, we take a closer look at \ARRAY{} and compare how the accessed memory size differs between schema widths $2$ and $16$: \CI{}~($16$$\rightarrow$$124$MiB), \UKI{}~($16$$\rightarrow$$16$MiB), and \UPI{}~($24$$\rightarrow$$132$MiB). Because during a traversal only the $16$MiB needs to be accessed, \UKI{} maintains its performance regardless of the schema width.

\begin{figure*}[h!]
    \centering
     \includegraphics[width=\linewidth]{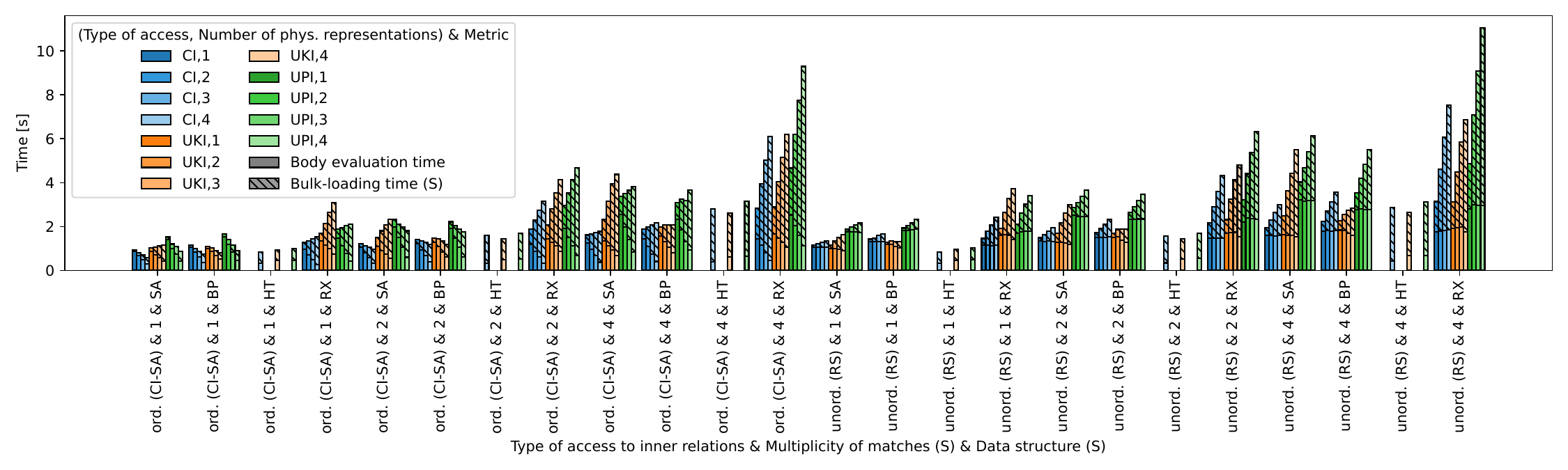}
     \vspace*{-0.7cm}
     \caption{Rule interweaving with four rules, where relation \smalltt{S} occurs in every rule and uses 1, 2, 3, or 4 physical representations.}
    \label{fig:rule_interweaving:query_and_insertion_time:without_hashtable}
 \vspace*{-0.5cm}
 \end{figure*}

\vspace*{-0.2cm}
\subsection{Sharing Physical Representations}
\label{ssec:share_factor}
\vspace*{-0.1cm}
Next, we analyze the case in which a single relation occurs \textit{multiple} times in the program. This can happen within the body of a single rule but also across rules. If, in this situation, the join key of one join participation is a prefix of the join key of another, an engine can either create a single physical representation on the longer join key and use it in both cases or create multiple representations specialized to the individual joins. While at first glance, the former sounds superior as it comes with less bulk-loading effort and a smaller memory footprint, it also has two potential downsides: First, even though one join key is a prefix of the order, this does not necessarily mean that both probe in the same order. Consequently, in the worst case, all but one join will experience unordered probes. Second, unordered data structures, such as \HT{}, cannot be used at all for joins that operate on a prefix of the index key, as they do not support range-lookups. 
To provoke and analyze this trade-off, we construct the following program which consists of four rules respectively bodies that all share the inner relation~\smalltt{S}:
\begin{figure}[h!]
    \centering
    \small
    \vspace*{-0.15cm}
    \begin{Verbatim}[commandchars=+\[\],xleftmargin=-0.45cm]
    (1) R_1(A), S(A,S_2,S_3,S_4,S_5,S_6,S_7,S_8)
    (2) R_2(B,A), S(A,B,S_3,S_4,S_5,S_6,S_7,S_8)
    (3) R_3(C,A,B), S(A,B,C,S_4,S_5,S_6,S_7,S_8)
    (4) R_4(D,B,C,A), S(A,B,C,D,S_5,S_6,S_7,S_8)
    \end{Verbatim}
    \vspace*{-0.15cm}
    \label{fig:rule_interweaving:rules}
    \vspace*{-0.2cm}
\end{figure}

\noindent Each body~$(i)$ has its exclusive outer relation~\smalltt{R\_i} which joins with the first $i$ attributes of~\smalltt{S}. Note that to provoke the previously mentioned ordering effect, we vary the order of the compound join key on the \smalltt{R}-side.
\begin{table}[h!]
    \centering
    \scriptsize
    \begin{tabular}{cC{1.8cm}C{1.5cm}C{1.5cm}C{1.5cm}}\toprule
         & \multicolumn{4}{c}{\textbf{Number of physical representations}}\\
         & \textbf{1} & \textbf{2} & \textbf{3} & \textbf{4} \\
         \midrule
        $\mathbf{R_1}$ & \textcolor{darkgreen}{$p_1(0\_1\_2\_3)$} & \textcolor{darkgreen}{$p_1(0\_1\_2\_3)$} & \textcolor{darkgreen}{$p_1(0\_1\_2\_3)$} & \textcolor{darkgreen}{$p_4(0)$} \\
        $\mathbf{R_2}$ & \textcolor{red}{$p_1(0\_1\_2\_3)$} & \textcolor{red}{$p_1(0\_1\_2\_3)$} & \textcolor{darkgreen}{$p_3(1\_0)$} & \textcolor{darkgreen}{$p_3(1\_0)$} \\
        $\mathbf{R_3}$ & \textcolor{red}{$p_1(0\_1\_2\_3)$} & \textcolor{darkgreen}{$p_2(2\_0\_1)$} & \textcolor{darkgreen}{$p_2(2\_0\_1)$} & \textcolor{darkgreen}{$p_2(2\_0\_1)$} \\
        $\mathbf{R_4}$ & \textcolor{red}{$p_1(0\_1\_2\_3)$} & \textcolor{red}{$p_1(0\_1\_2\_3)$} & \textcolor{red}{$p_1(0\_1\_2\_3)$} & \textcolor{darkgreen}{$p_1(3\_1\_2\_0)$} \\\bottomrule
    \end{tabular}
    \caption{The index key of each physical representation~$p_j$ created. A green entry means that the probes of $\mathbf{R_i}$ will happen in the same order as the physical representation.}
    \label{tab:rule_interweaving:physical_representations}
    \vspace*{-0.2cm}
\end{table}
In the experiment, we now vary the number of created physical representations for~\smalltt{S} from one to four. For each created physical representation that is shared, we chose its index key such that the probes of one join in which it participates are ordered. 
Table~\ref{tab:rule_interweaving:physical_representations} lists all created physical representations~$p_j(k_j)$ along with their index key~$k_j$, where $k_j=1\_0$ means that the index key is composed of attribute~$1$ followed by attribute~$0$. A green entry marks ordered probes, which increase with the number of specialized representations.
Figure~\ref{fig:rule_interweaving:query_and_insertion_time:without_hashtable} shows both the bulk-loading and body evaluation time where all relations \smalltt{R\_i} contains $1$M tuples while varying the number of physical representations, the type of physical representation of~\smalltt{S}, and the probe order caused by the individual  \smalltt{R\_i} relations as discussed before. To study the relationship to the additional bulk-loading time when creating more physical representations, we further vary the amount of duplicates per key (multiplicity) in \smalltt{S} from $1$~over~$2$~to~$4$, leading to $1$M, $2$M respectively $4$M tuples for~$S$.
First, we can see that for ordered probes, having more exclusive physical representations improves the overall evaluation time for a multiplicity of~$1$ and~$2$ (for all but \RX{}), even though the initialization effort increases, nicely exploiting the fitting index key. At the same time, we see in the case of a multiplicity of~$4$ that the additional bulk-loading effort can kill the advantage on the body evaluation side. Consequently, for ordered probes, the decision to create exclusive structures must consider the previously discussed relationship between body evaluation and bulk-loading effort, as well as potential restrictions due to a higher memory footprint. 
For unordered probes, creating exclusive physical representations does not pay off for the ordered structures \ARRAY{}, \BP{}, and \RX{} in any case, as the probe side does not utilize the available ordering. 
However, this can be handled by bringing in \HT{} for each exclusive index, which we also evaluated but do not show here due to space constraints. When following that strategy, increasing the number of exclusive \HT{} representations improves the evaluation time in all cases for a multiplicity of~$1$, and for \UPI{} for a multiplicity of~$2$. 
Regarding the memory consumption of the configurations with a multiplicity of~$4$, \CI{},1-\ARRAY{} has the lowest possible memory consumption with $246$MiB, which increases linearly to $984$MiB for \CI{},4-\ARRAY{}, nearly the same as the \CI{},4-\HT{} configuration requires. Using four indexes with \UPI{} consumes only $375$ MB across all data structures, except \RX{} ($612$ MB), highlighting \UPI{}'s use for sharing indexes to save memory.

\noindent
\setlength{\columnsep}{2pt}
\begin{figure*}[ht]
\begin{multicols}{4}
\removelatexerror{}
\footnotesize
\begin{algorithm}[H]
\DontPrintSemicolon
\SetAlgoLined
\label{alg:dedup1}
\caption{\small\DEDUPOne{}.}
\ForEach{\textnormal{tuple t}} {
    \If{\textnormal{!\textcolor{Orange}{\RbaseInCode{}.contains(t)}}} {
    \If{\textnormal{!\textcolor{Orange}{\RnewInCode{}.contains(t)}}} {
      \textnormal{\textcolor{Orchid}{\RnewInCode{}.append(t)}}
    }
  }
}
\textcolor{ForestGreen}{\RnewInCode{}.finished\_append()} \\
\textcolor{Orchid}{\RbaseInCode{}.append(\RnewInCode{})} \\
\textcolor{ForestGreen}{\RbaseInCode{}.finished\_append()}
\end{algorithm}
\vfill\null
\columnbreak
\begin{algorithm}[H]
\DontPrintSemicolon
\SetAlgoLined
\label{alg:dedup2}
\caption{\small\DEDUPTwo{}.}
\ForEach{\textnormal{tuple t}} {
  \If{\textnormal{!\textcolor{Orange}{\RnewInCode{}.contains(t)}}} {
    \If{\textnormal{!\textcolor{Orange}{\Rbase{}.contains(t)}}} {
      \textnormal{\textcolor{Orchid}{\RnewInCode{}.append(t)}}
    }
  }
}
\textcolor{ForestGreen}{\RnewInCode{}.finished\_append()} \\
\textcolor{Orchid}{\RbaseInCode{}.append(\RnewInCode{})} \\
\textcolor{ForestGreen}{\RbaseInCode{}.finished\_append()}
\end{algorithm}
\vfill\null
\columnbreak
\begin{algorithm}[H]
\DontPrintSemicolon
\caption{\DEDUPThree{}.}
\label{alg:dedup3}
\ForEach{\textnormal{tuple t}} {
  \If{\textnormal{!\textcolor{Orange}{\RbaseInCode{}.contains(t)}}} {
    \textnormal{\textcolor{Orchid}{\RbaseInCode{}.append(t)}} \\
    \textnormal{\textcolor{Orchid}{\RnewInCode{}.append(t)}}
  }
}
\textcolor{ForestGreen}{\RbaseInCode{}.finished\_append()} \\
\textcolor{ForestGreen}{\RnewInCode{}.finished\_append()}
\end{algorithm}
\vfill\null
\columnbreak
\begin{algorithm}[H]
\caption{\DEDUPFour{}.}
\label{alg:dedup4}
\ForEach{\textnormal{tuple t}} {
  \If{\textnormal{!\textcolor{Orange}{\RnewInCode{}.contains(t)}}} {
    \textnormal{\textcolor{Orchid}{\RnewInCode{}.append(t)}}
  }
}
\textcolor{ForestGreen}{\RnewInCode{}.finished\_append()} \\
\ForEach{\textnormal{tuple t$\in$\RbaseInCode{}}} {
  \If{\textnormal{\textcolor{Orange}{\RnewInCode{}.contains(t)}}} {
     \textcolor{Periwinkle}{\RnewInCode{}.remove(t)}
  }
}
\textcolor{Orchid}{\RbaseInCode{}.append(\RnewInCode{})} \\
\textcolor{ForestGreen}{\RbaseInCode{}.finished\_append()}
\end{algorithm}
\end{multicols}
\vspace*{-0.3cm}
\caption{Four strategies to build-up the recursive relation. The first variant~\DEDUPOne{} was already shown in Algorithm~\ref{alg:physical_execution_in_souffle}.}
\vspace*{-0.4cm}
\end{figure*}
\noindent

\vspace*{-0.3cm}
\subsection{Build-up of the Recursive Relation}
\label{ssec:deduplication}
\vspace*{-0.1cm}

Let us now extend the discussion to the aspects that are specific to the recursive evaluation, where a relation is incrementally built up with new facts.
To do so, we focus on a single iteration of the semi-naive evaluation and study how found facts are identified as new and unique and if so, subsequently appended to the recursive relation. To take the already analyzed body evaluation out of the picture, we test the build-up of the recursive relation on a set of pre-generated tuples with certain characteristics that represent the tuples an actual body evaluation would have found. As there are different strategies to identify and append new and unique tuples, which vary in the order in which operations are executed, we evaluate four different options: \\
\indent Strategy~\DEDUPOne{} (Algorithm~\ref{alg:dedup1}) resembles the already seen variant implemented by Soufflé. As an alternative, we evaluate~\DEDUPTwo{} (Algorithm~\ref{alg:dedup2}) where the two \smalltt{contains()} checks are reversed in order. This might have a positive effect if many duplicates are generated, as per duplicate group, \DEDUPTwo{} requires only one containment check into the base relation.
In strategy~\DEDUPThree{} (Algorithm~\ref{alg:dedup3}), we do not decouple the identification of new tuples from the merging of them into the base relation, but rather merge each new tuple into the base relation as early as possible. This avoids a separate pass over the new relation; however, it also leads to unordered \smalltt{append()} operations into the base relation, if the found tuples arrive unordered.  
Strategy~\DEDUPFour{} (Algorithm~\ref{alg:dedup4}) eliminates the containment checks in the base relation. Instead, after building up the new relation, it iterates over the base relation and removes all already contained entries from the new relation. After that, it merges the new relation into the base relation.  
In terms of physical representations, we stick to \CI{} as access type to keep the analysis readable and vary  the more relevant data structure, where the chosen physical representation is used for \Rbase{}, \Rdelta{}, and \Rnew{}.
\begin{figure}[h!]
    \centering
     \begin{subfigure}{.49\columnwidth}
        \centering
         \includegraphics[width=\linewidth]{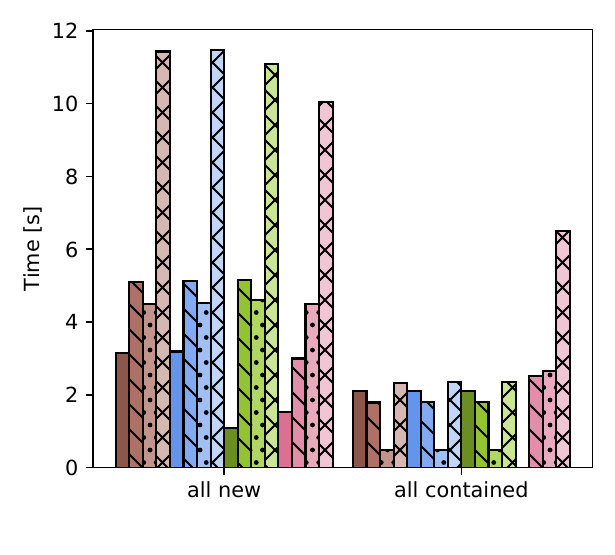}
         \vspace*{-0.8cm}
         \caption{All unique, unordered arrival.}
         \label{fig:exp:impact_of_deduplication_strategy:no_duplicates_unordered}
     \end{subfigure}
     \begin{subfigure}{.49\columnwidth}
        \centering
         \includegraphics[width=\linewidth]{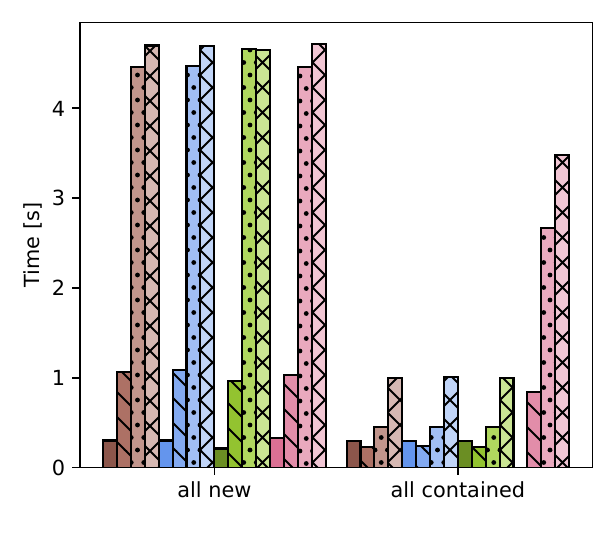}
         \vspace*{-0.8cm}
         \caption{All unique, ordered arrival.}
         \label{fig:exp:impact_of_deduplication_strategy:no_duplicates_ordered}
     \end{subfigure}
     
    \begin{subfigure}{.49\columnwidth}
        \centering
         \includegraphics[width=\linewidth]{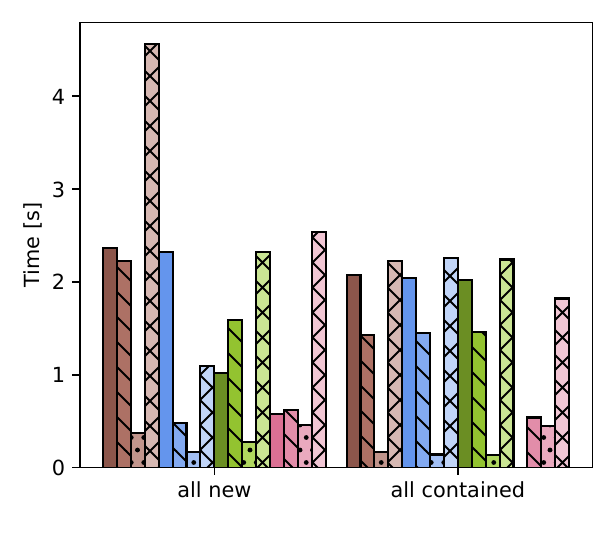}
         \vspace*{-0.8cm}
         \caption{All 100$\times$, unordered arrival.}
         \label{fig:exp:impact_of_deduplication_strategy:with_duplicates_unordered}
     \end{subfigure}
     \begin{subfigure}{.49\columnwidth}
        \centering
         \includegraphics[width=\linewidth]{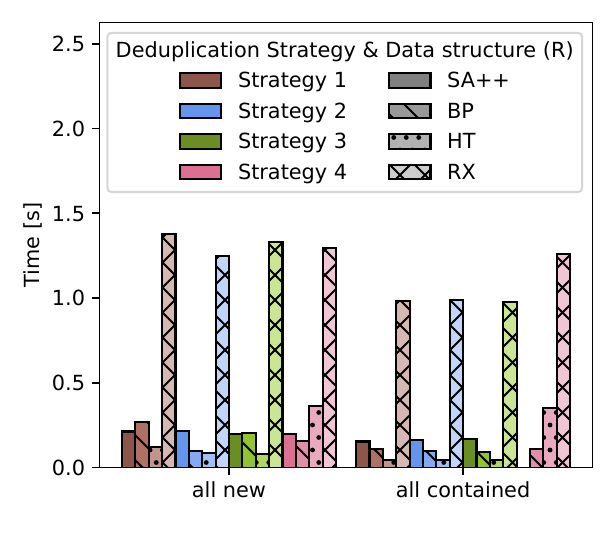}
         \vspace*{-0.8cm}
         \caption{All 100$\times$, ordered arrival.}
         \label{fig:exp:impact_of_deduplication_strategy:with_duplicates_ordered}
     \end{subfigure}
     
     \caption{Build-up of recursive relation for different strategies.}
     \label{fig:exp:impact_of_deduplication_strategy}
     \vspace*{-0.5cm}
\end{figure}
We set up a recursive relation with two attributes and fill its base relation~\Rbase{} with $6.5$M unique tuples ($100$MiB in total). We then generate the same amount of found tuples, where we test the two cases in which either all of these tuples are new to the base relation or all are already contained. Additionally, we vary the number of found tuples that are duplicates. As a third dimension, we distinguish whether the found tuples arrive unordered or ordered with respect to the order maintained by \Rbase{}, \Rdelta{}, and \Rnew{}. \\
\indent Figure~\ref{fig:exp:impact_of_deduplication_strategy} shows the results. From a distance, we can already make two interesting observations: 
First, the strategy makes a significant performance difference in individual cases, and different strategies win in different situations, so we should consider multiple strategies in the following. However, we can directly rule out~\DEDUPFour{}, as it did not outperform the other strategies for any setup. This simplifies the analysis, as we no longer need to consider the performance of \smalltt{remove()}.   
Second, the combination of workload characteristics and strategy impacts which physical representation performs best. Let us look at this in detail, starting with the case where all found tuples are unique. 
Figure~\ref{fig:exp:impact_of_deduplication_strategy:no_duplicates_unordered} shows it for an unordered arrival of found tuples. As both \DEDUPOne{} and \DEDUPTwo{} perform for each tuple unordered \smalltt{contains()} operations on \Rbase{} and \Rnew{} followed by unordered \smalltt{append()} operations to \Rnew{} respectively ordered \smalltt{append()} operations to \Rbase{}, they have the same performance. \DEDUPThree{} has to execute fewer \smalltt{contains()} operations but buys this by performing only unordered \smalltt{append()} operations. This is exploited by \HT{}, but even more so by \ARRAYALTERNATIVE{}, which sorts all appended tuples in a batch when \smalltt{append\_finished()} happens, clearly outperforming the alternatives. This leads to an improvement over the worst choice by almost $11\times$.  
If all found tuples are already contained \Rbase{}, the picture is entirely different, as all strategies now have to perform only unordered \smalltt{contains()} operations. Although \DEDUPOne{} and \DEDUPThree{} have to perform fewer of these, all strategies perform very similarly, where \HT{} outperforms the remaining physical representations as it handles unordered \smalltt{contains()} the best. 
If tuples arrive in order, as shown in Figure~\ref{fig:exp:impact_of_deduplication_strategy:no_duplicates_ordered}, all ordered physical representations perform significantly faster. While all of them can handle the ordered \smalltt{contains()} operations very well, \ARRAYALTERNATIVE{} clearly wins over \BP{} due to its superior \smalltt{append()} performance. \HT{} and \RX{} are by far the worst choices in this situation.
If all in order arriving tuples are already contained, the performance of all methods further improves over the corresponding unordered case. However, the previously winning \HT{} is now naturally a bad choice, while \BP{} outperforms the remaining ordered physical representations, showing its superior \smalltt{contains()} performance.   
When having duplicates in the found tuples, the picture changes again significantly, as the number of duplicates not only impacts the balance between the occurring \smalltt{contains()} and \smalltt{append()} operations, but also the effort of the \smalltt{append\_finished()} operation for \ARRAYALTERNATIVE{}. 
Figure~\ref{fig:exp:impact_of_deduplication_strategy:with_duplicates_unordered} shows the case when all found tuples are new. For \DEDUPTwo{} and \DEDUPThree{}, which are naturally the best strategies when handling duplicates, \ARRAYALTERNATIVE{} is no longer the best, but the worst option, as its duplicate removal operation is expensive. Consequently, \HT{} is around $5\times$ faster than \ARRAYALTERNATIVE{} in this case. 
In contrast, when the found tuples are already contained, duplicates have basically no effect.   
Finally, let us look at the ordered case for duplicates in Figure~\ref{fig:exp:impact_of_deduplication_strategy:with_duplicates_ordered}. Interestingly, when facing new tuples, \HT{} still outperforms the ordered physical representations. 
Overall, these findings show how carefully the physical representations for the recursive relation must be selected.

\vspace*{-0.2cm}
\section{\mbox{Automatic Physical Representation Selection}}
\label{sec:automatic_selection}
\vspace*{-0.1cm}

In Section~\ref{sec:non_recursive_evaluation}, we carefully analyzed the impact of various workload characteristics on the choice of physical representation. We will now materialize the insights gained in a selection mechanism that can identify suitable physical representations for a given workload. Precisely, for a workload, we first build what we call a workload signature, which captures all relevant characteristics of the workload in a condensed form. Using this signature, which captures the number of performed operations and the potential for aligning them, we compute a suitable index key for each physical representation that determines the actual alignment of operations. With that, we use three decision trees to select the data structures, access types, and amount of sharing for the entire workload.

To demonstrate the effectiveness of our selection mechanism, we apply it in the following on four real-world workloads from the domain of program analysis and graph analytics to identify effective configurations.
Table~\ref{tab:datalog_programs_and_datasets} provides an overview of the workloads and states from. For our running example~\AndersensAnalysis{}, it looks as follows:

\noindent
\begin{scriptsize}
\begin{Verbatim}
points_to(y,x):-address_of(y,x)
points_to(y,x):-assign(y,z),points_to(z,x)
points_to(y,w):-load(y,x),points_to(x,z),points_to(z,w)
points_to(y,w):-store(y,x),points_to(y,z),points_to(x,w)
\end{Verbatim}
\end{scriptsize}

\begin{table}[h!]
\centering
\begin{tabular}{cccc}\toprule
\textbf{Datalog program} & \textbf{Dataset} & \textbf{Vertices} & \textbf{Edges} \\\midrule
\cellcolor{gray!35}
\textbf{Program analysis} &&&\\
Andersen's analysis\cite{phd:andersenAnalysis, paper:recstep} & 50000\cite{paper:recstep} & - & -  \\
\cellcolor{gray!35}
\textbf{Graph analytics} &&&\\
Reachability\cite{paper:bigdatalog, paper:recstep, paper:RaSQL} & livej. \cite{online:livejournal, paper:livejournalIntroducing} & 4,847k & 68,993k \\
Same generation\cite{paper:bigdatalog, paper:recstep} & G10K-0.001 \cite{online:GTgraphGenerator} & 10k & 100k \\
Transitive closure\cite{paper:bigdatalog, paper:recstep} & G10K-0.01 \cite{online:GTgraphGenerator} & 10k & 1,000k \\\bottomrule
\end{tabular}
\caption{Real-world workloads.}
\label{tab:datalog_programs_and_datasets}
\vspace{-0.5cm}
\end{table}

\vspace*{-0.2cm}
\subsection{Workload Signatures}
\vspace*{-0.1cm}

At the example of \AndersensAnalysis{}, let us see how the workload signature looks like.

\begin{figure}[h!]
     \begin{subfigure}{0.99\columnwidth}
        \centering
\large
\scalebox{0.45}{%
    \begin{tikzpicture}[node distance=5cm]
    \begin{scope}[node distance=3cm]
    \node[text=gray]         (G) {};
    \node[text=loadColor]           (A) [right of=G] {load$(y,x)$};
    \end{scope}
    \node[text=pointToDeltaColor]   (B) [right of=A] {$\text{points\_to}_{\text{delta}}(x,z)$};
    \begin{scope}[node distance=0.35cm]
    \node[text=pointToDeltaColor]   (B2) [above of= B] {\marvosymLightning};
    \end{scope}
    
    \node[text=pointToColor]        (C) [right of=B] {$\text{points\_to}(z,w)$};
    \begin{scope}[node distance=1.5cm]
    \node[]        (E) [above of= C] {};
    \node[]        (F) [below of= C] {};
    \end{scope}
    \node[text=pointToColor]        (E2) [right of= E] {$\text{points\_to}(y,w)$};
    \node[text=pointToNewColor]        (F2) [right of= F] {$\text{points\_to}_{\text{new}}(y,w)$};
    
    \draw[ ->, >=stealth] (G) -- node[midway, above] {$\text{bulk\_insert()}$} (A);
    \draw[ ->, >=stealth] (G) -- node[midway, below, text=gray] {1.6M} (A);
    \draw[->, >=stealth, loop above] (A) to node[midway, above, align=center, text=gray] {iter$()$/returned\\15/24.8M} (A);
    \draw[ ->, >=stealth] (A) -- node[midway, above] {probe$(x)$/returned} (B);
    \draw[ ->, >=stealth] (A) -- node[midway, below, text=gray] {24.8M/135.3K} (B);
    \draw[dashed, ->, >=stealth] (B) -- node[midway, above] {probe$(z)$/returned} (C);
    \draw[dashed, ->, >=stealth] (B) -- node[midway, below, text=gray] {135.3K/10.0K} (C);
    
    \draw[dotted, ->, >=stealth] (C) -- node[midway, left] {w} (E);
    \draw[dotted, ->, >=stealth] (C) -- node[midway, left] {w} (F);

    \draw[dashed, ->, >=stealth] (E) -- node[midway, above] {contains$(y,w)$} (E2);
    \draw[dashed, ->, >=stealth] (E) -- node[midway, below, text=gray] {10.0K} (E2);

    \draw[dashed, ->, >=stealth] (F) -- node[midway, above, align=center] {contains$(y,w)$\\append$(y,w)$} (F2);
    \draw[dashed, ->, >=stealth] (F) -- node[midway, below, align=center, text=gray] {10.0K\\9.9K} (F2);

    \draw[bend left=22, ->, >=stealth] (A) to node[midway, above] {y} (E); 
    \draw[bend left=-22, ->, >=stealth] (A) to node[midway, above] {y} (F); 
    \end{tikzpicture}
}
         \vdots
         \caption{Alignment and number of operations (shown for one rule).}
        \label{fig:real_world:andersen_analysis:operations_plot}
     \end{subfigure}

    \begin{subfigure}{0.49\columnwidth}
        \centering
         \includegraphics[width=\linewidth]{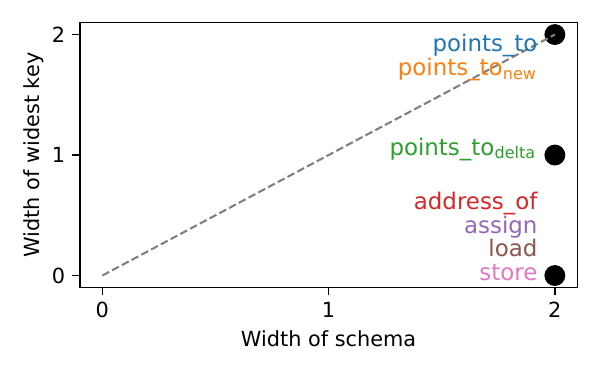}
         \caption{Schema width vs key width.}
         \label{fig:real_world:andersen_analysis:schema_vs_key}
     \end{subfigure}
     \begin{subfigure}{0.49\columnwidth}
        \centering
         \includegraphics[width=\linewidth]{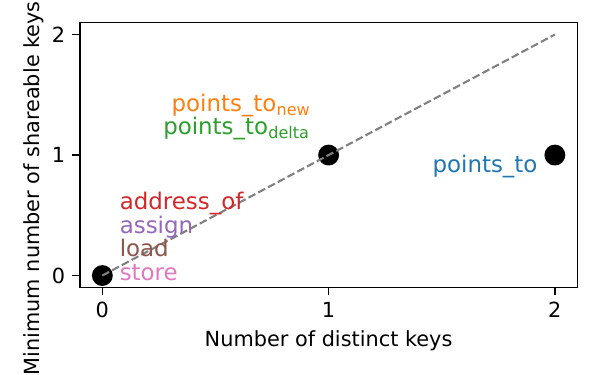}
         \caption{Distinct vs shareable keys.}
         \label{fig:real_world:andersen_analysis:distinct_vs_sharing}
     \end{subfigure}
     \vspace*{-0.1cm}
        \caption{\mbox{Signature of \AndersensAnalysis{}.}}
        \label{fig:real_world:andersen_analysis}
        \vspace*{-0.5cm}
\end{figure}

\textbf{Alignment and number of operations} (Figure~\ref{fig:real_world:andersen_analysis:operations_plot}). Across all experiments, we have observed that the performance is highly influenced by two properties: (1)~Whether operations exploit an alignment or not and (2)~the overall mixture of operations.
Regarding~(1), we capture this ``alignment'' as follows: First, for each body evaluation of the program, starting at the iteration over the outermost relation and ending with the build-up of the head relation, we create an alignment graph. This alignment graph captures, for each level of the nested loop, whether an alignment between the current relation and the relations on the next level is possible. This effectively creates alignment paths down the nested loops. 
Note that paths of alignment involving more than one edge can occur only if the EDBs contain aligned attributes.  
Regarding (2)~, we equip the previously created alignment graph with the actual  mixture of carried out operations, such as \smalltt{bulk\_load()}, \smalltt{contains()}, \smalltt{append()}, and \smalltt{probe()} operations, allowing to determine the importance of each operation for the overall evaluation. To capture this aspect of the workload in a digestible way, we instrument the evaluation code and perform a profiling run.

Figure~\ref{fig:real_world:andersen_analysis:operations_plot} shows one out of the eight generated graphs for \AndersensAnalysis{}.
In the alignment graph of the rule \smalltt{points\_to(y,w) :- load(y,x), points\_to(x,z), points\_to(z,w)}, the \smalltt{load} relation is potentially fully aligned with the \smalltt{points\_to\textsubscript{delta}} relation during the probe (solid line), but also partially aligned with \smalltt{points\_to} and \smalltt{points\_to\textsubscript{new}} during \smalltt{contains()/append()} (dashed line). Consequently, the choice of physical representation for \smalltt{load} has a strong impact on the choice of physical representations for the other relations. Note that although from \smalltt{points\_to\textsubscript{delta}} the probing continues into \smalltt{points\_to}, this access can never be aligned (dotted line), as there is no inherent alignment between the two attributes of \smalltt{points\_to\textsubscript{delta}}. If the dataset contained such an alignment, the impact of the \smalltt{load} relation would be even stronger.
When looking at the amount of performed operations, we can see that the \smalltt{probe()} operations into \smalltt{points\_to\textsubscript{delta}} significantly dominate the remaining operations, signaling its importance for the choice of physical representation.

\textbf{Schema and key width vs number of distinct and shareable keys} (Figure~\ref{fig:real_world:andersen_analysis:schema_vs_key} and Figure~\ref{fig:real_world:andersen_analysis:distinct_vs_sharing}).
In the experiments, we have seen that the decision about physical representation is also highly influenced by the widths of relations and keys, the total number of occurrences of these in the program, and the potential to share physical representations. To capture these aspects, parts three and four of the signature connect for each relation the following properties: The schema width and the width of the widest occurring key, as well as the number of distinct keys and the minimum number of shareable keys. 

\vspace*{-0.3cm}
\subsection{Selection Mechanism}
\vspace*{-0.2cm}

Let us now see how our selection mechanism utilizes the signature to come up with an efficient selection.

\textbf{Index keys}. As we have observed the importance of aligned accesses, the index keys must be computed first, as they determine where aligned accesses can happen (if ordered data structures are chosen later). If multiple different alignment paths exist, as in our example graph, we want to select the path that is likely to benefit the most from aligned accesses. To identify this path and the index key selection leading to it, we start at the left-most relation in the body (\smalltt{load(y,x)} in our case) and follow the alignment path when the index key is set to the first attribute only and when it is set to the second attribute only, while we compute the total number of performed operations on each path. We then select the index key of the path that performs the most operations. In our case, an index key on the second attribute is selected, as it leads to $24.8$M ordered \smalltt{probe()} operations, where an index key on the first attribute leads only to $29$K (partially) ordered accesses into \smalltt{points\_to} and \smalltt{points\_to\textsubscript{new}}. The index keys of all following relations are then set such that the ordering on the most expensive path is kept intact as long as possible. Overall, for \AndersensAnalysis{}, this results in the index keys \smalltt{address\_of(0\_1)}, \smalltt{assign(1)}, \smalltt{load(1)}, \smalltt{store(0)}, \smalltt{points\_to\textsubscript{delta/new}(0\_1)}. Note that at this point, both \smalltt{points\_to(0)} and \smalltt{points\_to(0\_1)} will be considered and passed on to the next stage, as they allow for sharing.

\textbf{Decision trees}. After computing all index keys, our selection mechanism next has to determine the physical representations and their amount of sharing under multiple occurrences. To do so, it incrementally uses three decision trees applied at the rule level that materialize the previously identified most important findings into practical decisions aimed at balancing performance and space efficiency.

\begin{figure}[h!]
    \centering
     \begin{subfigure}{\columnwidth}
        \centering
         \includegraphics[page=2, width=.85\linewidth, trim={0 48cm 37.5cm 0}, clip]{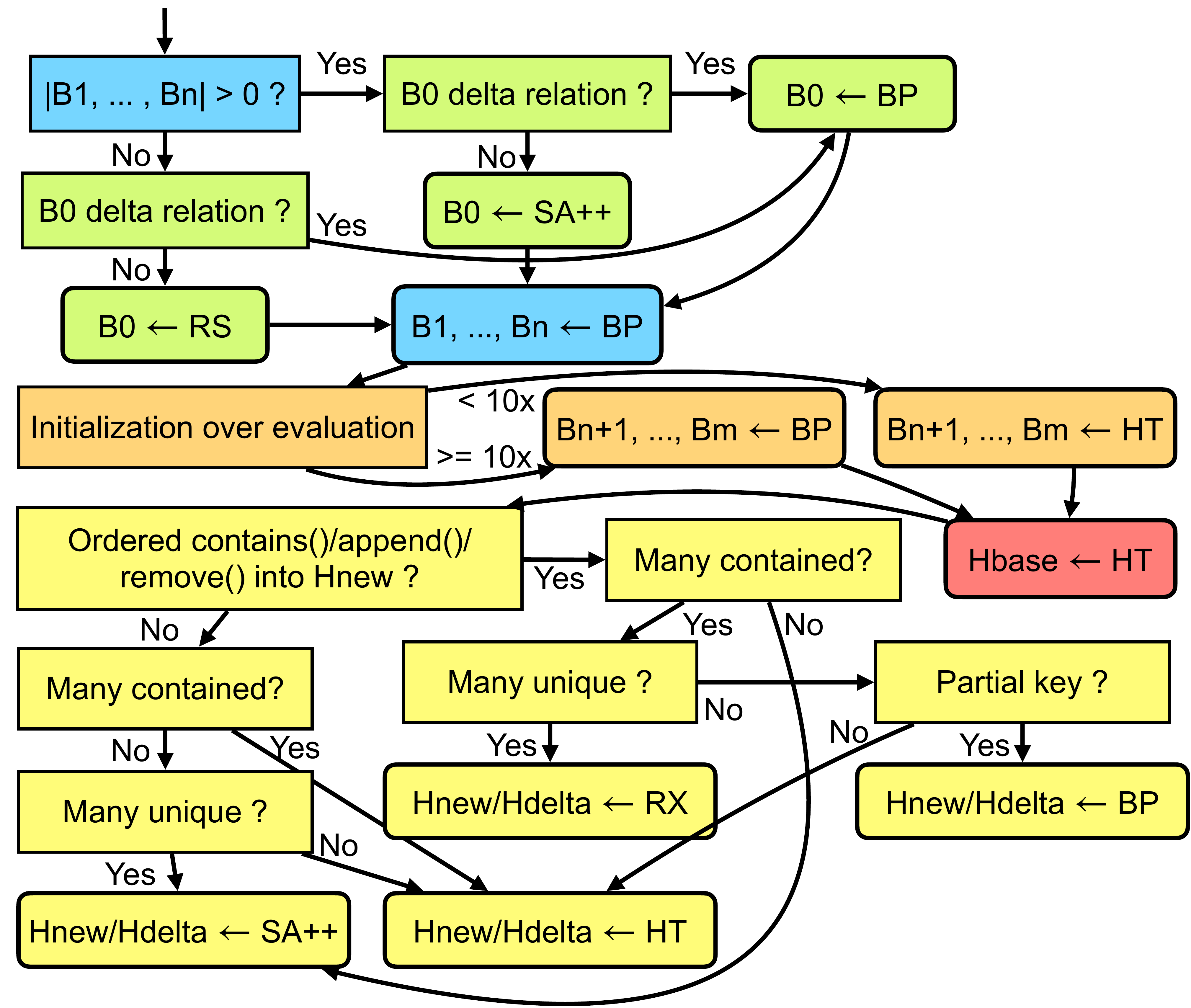}
         \caption{Selecting data structures.}\label{fig:decision_trees:data_structures}
     \end{subfigure}
     \begin{subfigure}{\columnwidth}
        \centering
         \includegraphics[page=3, width=.85\linewidth, trim={0 80.5cm 37.5cm 0}, clip]{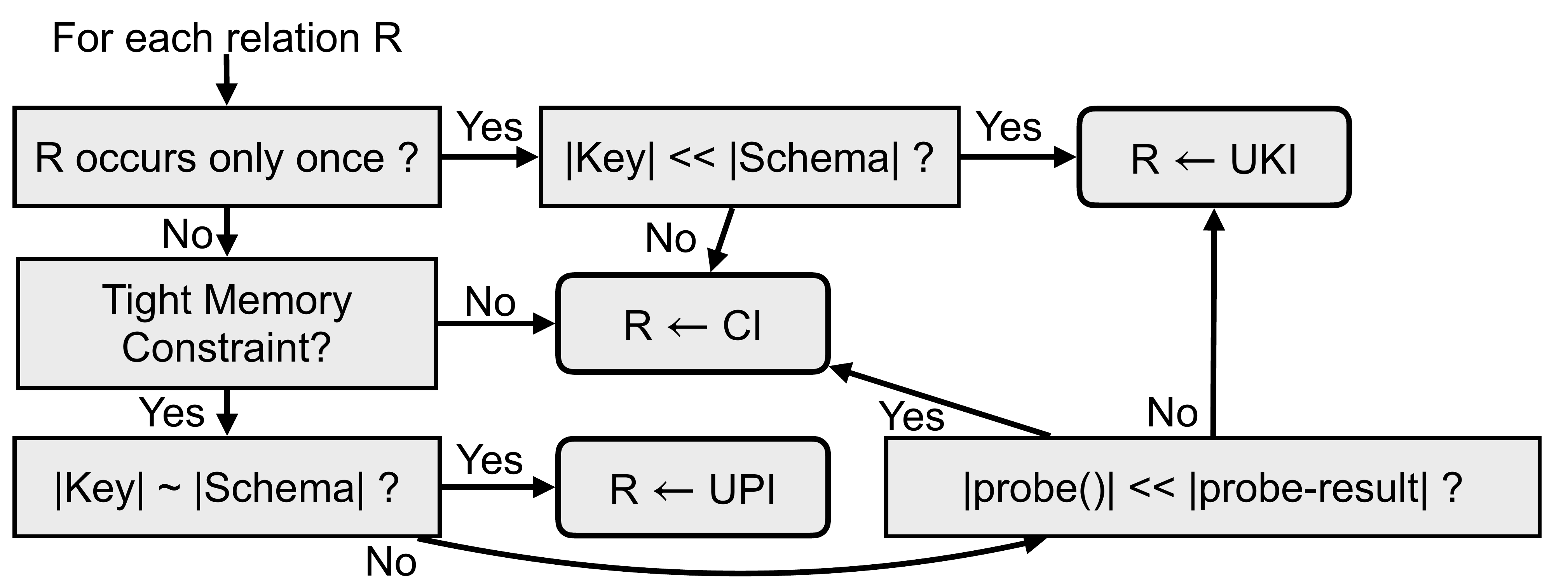}
         \caption{Selecting access types.}\label{fig:decision_trees:access_types}
     \end{subfigure}

     \begin{subfigure}{\columnwidth}
        \centering
         \includegraphics[page=4, width=.85\linewidth, trim={0 74cm 37.5cm 0}, clip]{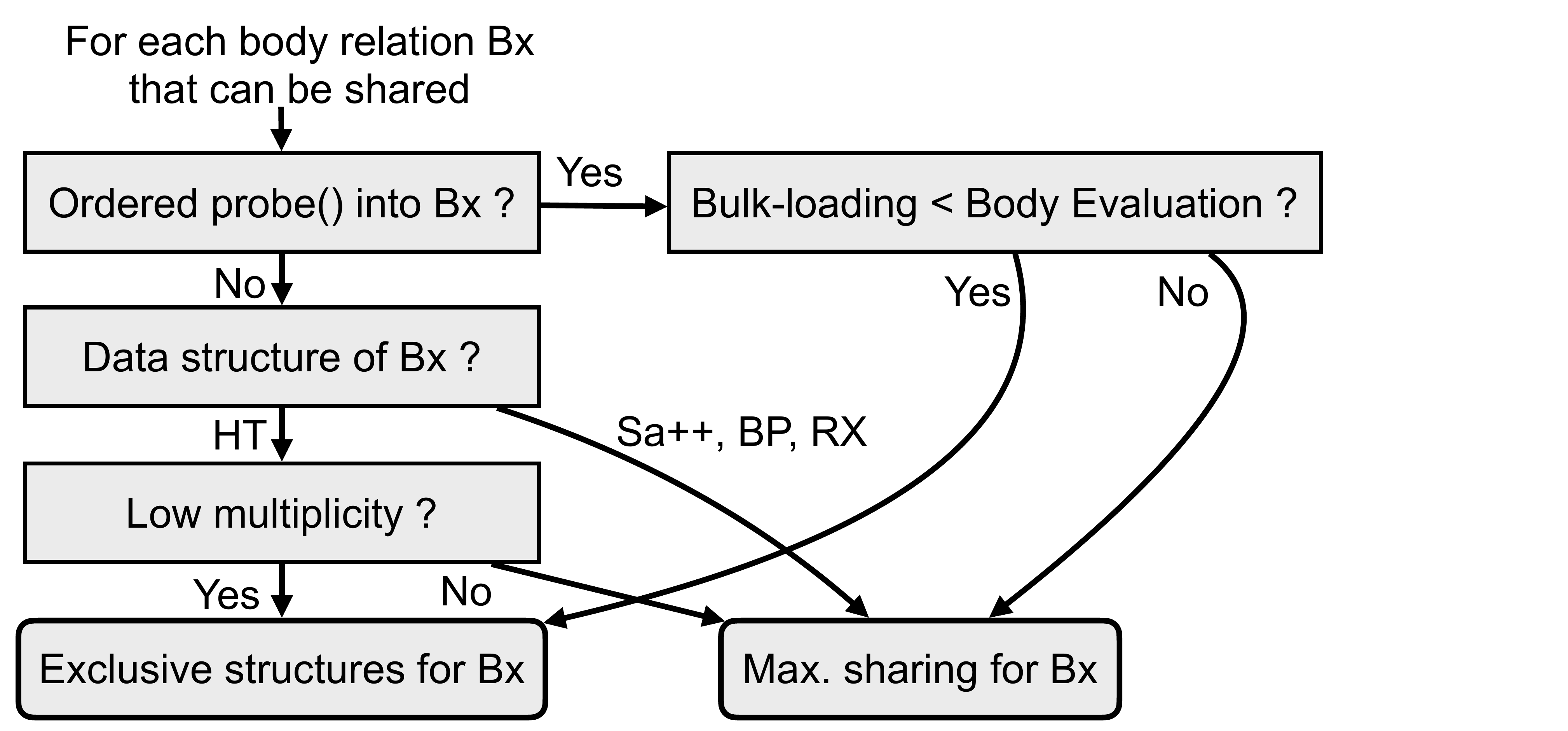}
         \caption{Selecting shared physical representations.}\label{fig:decision_trees:share_factor}
     \end{subfigure}
    
     \caption{Decision trees. $B_0$ is the first body relation, $B_1$...$B_n$ exploit alignment, $B_{n+1}$...$B_m$ cannot exploit alignment.}
     \label{fig:decision_trees}
     \vspace*{-0.5cm}
\end{figure}

To select the data structure, we first split the rule to instantiate into different parts. Precisely, in the body, we differentiate between the outer-most relation~($B_0$), the relations that can exploit alignment~($B_1$ to $B_n$), and the relations that cannot exploit alignment~($B_{n+1}$ to $B_m$) with $B_0$. In the head, we differentiate between $H_{base}$ and $H_{new}$. The decision tree shown in Figure~\ref{fig:decision_trees:data_structures} then incrementally instantiates each individual part of the rule based on a sequence of criteria that (a)~aim at creating aligned accesses, (b)~factor in the ratio of initialization and evaluation, and (c)~the characteristics of the generated tuples. Note that conflicting decisions for $H_{new}$ and $H_{delta}$ are both dragged into the next decision tree, where they might result in the creation of both representations or the sharing of a single one, if possible and recommended.
For \AndersensAnalysis{}, applying this decision tree results in the selection of \ARRAY{} for the $B_0$ EDBs \smalltt{address\_of}, \smalltt{assign}, \smalltt{load}, and \smalltt{store} to kick off paths of aligned accesses, the selection of \HT{} for \smalltt{points\_to} to efficiently handle the \smalltt{contains()} operations into the base relation, and the selection of~\BP{} for \smalltt{points\_to\textsubscript{delta}} to efficiently handle ordered probes.
After selecting the data structure, the access type is determined next. Here, we clearly prioritize \CI{}, unless the key is significantly narrower than the schema or a tight space constraint exists. For \AndersensAnalysis{}, we do not assume that such a one exists. Consequently, \CI{} is selected for all physical representations.
The final decision tree decides for each physical representation that can be shared whether it actually should be shared or not. Again, we factor in whether ordered probes occur into the relation and the ratio between initialization effort and evaluation, but also the chosen data structure and the multiplicity when probing, if that information is available. As a result, we either choose to create exclusive physical representations for the relation or to share as much as possible. For \AndersensAnalysis{}, where this decision tree is only applied for \smalltt{points\_to}, it decides to create the two exclusive physical representations \smalltt{points\_to(0)} and \smalltt{points\_to(0\_1)} (both \CI{}-\HT{}).

\subsection{Experimental Results and Baseline Comparison}

\begin{figure}[h!]
    \centering
     \begin{subfigure}{.38\columnwidth}
        \centering
         \includegraphics[width=\linewidth]{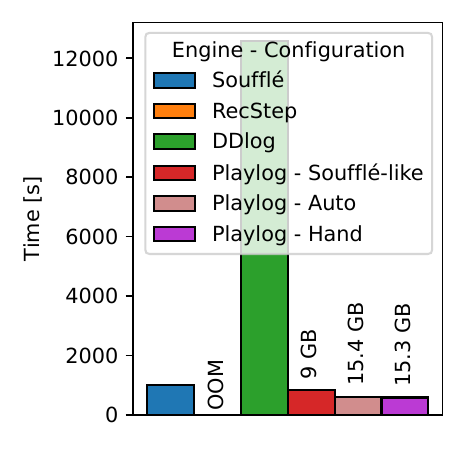}
         \vspace*{-0.8cm}
         \caption{Transitive closure.}
         \label{fig:exp:engine_comparison:transitive_closure}
     \end{subfigure}
     \begin{subfigure}{.38\columnwidth}
        \centering
         \includegraphics[width=\linewidth]{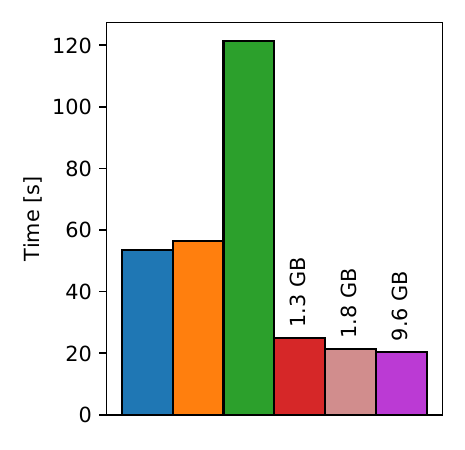}
         \vspace*{-0.8cm}
         \caption{Reachability.}
         \label{fig:exp:engine_comparison:reach}
     \end{subfigure}
    \begin{subfigure}{.38\columnwidth}
        \centering
         \includegraphics[width=\linewidth]{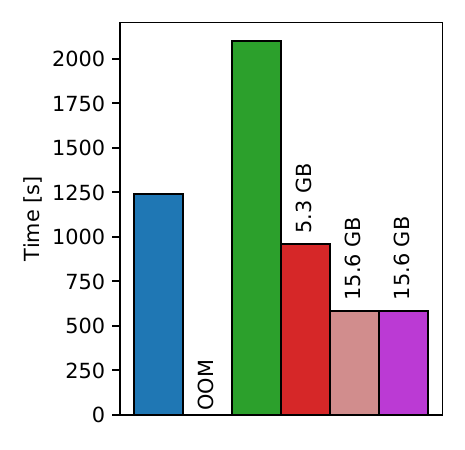}
         \vspace*{-0.8cm}
         \caption{Same generation.}
         \label{fig:exp:engine_comparison:same_generation}
     \end{subfigure}
    \begin{subfigure}{.38\columnwidth}
	\centering
             \includegraphics[width=\linewidth]{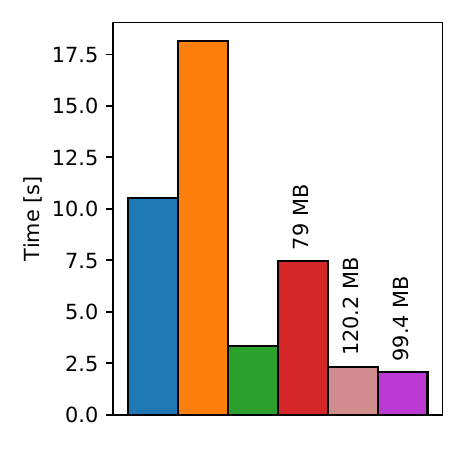}
	       \vspace*{-0.8cm}
         \caption{Andersen analysis.}
         \label{fig:exp:engine_comparison:andersen_analysis}
     \end{subfigure}
     \caption{Comparison of generated  configurations.}
     \label{fig:real_world_workloads:signature_optimized}
     \vspace*{-0.5cm}
\end{figure}

To assess the quality of our selection mechanism, we perform the following experiment: For each of the four real-world workloads, we apply the selection mechanism to generate a configuration, referred to as \engine{}-Auto. We then compare this configuration to a Soufflé-like configuration in our engine, where only \CI{}-\BP{} \blue{with Soufflé's index keys} is used for all physical representations, referred to as \engine{}-Soufflé-like, as well as the best hand-tailored configuration referred to as \engine{}-Hand.
Additionally, to put the results of \engine{} into perspective, we include the time to evaluate the workloads in three other comparable single-node main-memory engines, namely Soufflé, RecStep, and DDLog.
Figure~\ref{fig:real_world_workloads:signature_optimized} shows the results for all four workloads. We can observe that, despite their apparent simplicity, there are vast performance differences observable. 
Let us first look at the results from \engine{}. We can see that for all four workloads, the Soufflé-like configuration performs the worst, while our hand-tailored configuration shows the best performance, ranging between a speedup from $1.2$x~(Reachability) to $3.5$x~(\AndersensAnalysis{}). This shows that the selection of appropriate physical representations actually matters for real-world workloads as well. Further, we can see that for all workloads, our selection mechanism is able to identify a configuration that is close to the hand-tailored one, being at most $1.1$x~slower.
The baseline systems are never able to beat the automatic selection of \engine{}, where RecStep even runs out of memory for Same generation and Transitive closure due to the materialization of large intermediate join results. Apart from that, during an iteration of the semi-naive evaluation, RecStep does not perform containment checks on the \textit{new} or \textit{base} relation, resulting in many duplicates and a costly deduplication and set-difference operation.
DDLog is designed for incremental computation and maintains temporal indexes (indexed by logical time) that contain previous versions of relations. The evaluation performed in Differential dataflow (DDLog's backend) is not equivalent to the left-deep join tree used by Soufflé and \engine{}. This can result in more frequent recursive rule evaluation, which was the case for $\text{same\_generation}$ and $\text{transitive\_closure}$. But can also increase the performance for relations defined by many rules $\text{points\_to}$ in the \AndersensAnalysis{}.

\vspace*{-0.2cm}
\section{Related Work}
\label{sec:related_work}
\vspace*{-0.1cm}
\cite{paper:survey} provides an excellent overview over the current Datalog landscape. 
Regarding physical representations, other works have also studied the impact of different options: In~\cite{paper:souffle:btree, paper:souffle:trie}, the authors compare a concurrent BTree with a concurrent radix tree in the context of Soufflé. In~\cite{paper:recstep}, which proposes RecStep that builds on the relation system QuickStep~\cite{paper:quickStep}, the authors mainly use a hash table, but also implement a bit-matrix to handle dense graph datasets well. The relatively old system BDDBDDB~\cite{paper:bddbddb} uses exotic binary decision diagrams. BigDatalog builds on Apache Spark\cite{paper:ApacheSpark} with specialized RDDs, whereas SociaLite introduced tail-nested tables, which are pretty similar to radix trees and space efficient for social graphs. The fact that different systems took very different routes in this regard shows that our study targets an important problem, especially since representations were never evaluated under a common hood. 
Regarding query optimization in Datalog, other sub-problems have been addressed in related work: In~\cite{paper:automaticIndexSelection}, a technique has been proposed to automatically select the minimal amount of required indexes by sharing physical representations which potentially reduces the memory consumption. However, as shown in subsection~\ref{ssec:share_factor}, this might prevent the benefits. In~\cite{paper:souffleJoinOptimizer}, the authors propose a join ordering optimizer for Soufflé. To do so, the authors extend Soufflé by an offline profiling pass that runs the program in its default order and collects statistics on a representative dataset that must be provided additionally. Based on these statistics, an optimal join order is identified taking the different recursiveness across the iterations into account. The findings of our work are orthogonal to the techniques presented therein.

\vspace*{-0.2cm}
\section{Conclusion}
\label{sec:conclusion}
\vspace*{-0.1cm}

In this work, we carefully analyzed the special requirements of Datalog regarding the selection of different physical representations. In our highly flexible Datalog engine~\engine{}, we first studied the impact of various workload characteristics on the individual physical representations. Based on our findings, we derived decision trees and an automatic selection mechanism that is able to identify high-performance configurations for real-world workloads that are close to hand-optimized counterparts and to full-fledged Datalog systems.

\noindent \textbf{Acknowledgments}. This research is funded by the DFG (German Research Foundation) – project 508316729.

\section*{AI-Generated Content Acknowledgement}

GitHub Copilot assisted in some of the boilerplate code generation during development.
\bibliographystyle{IEEEtran}
\bibliography{datalog}

\end{document}